\documentclass[iop,useAMS]{emulateapj}
\usepackage[fleqn]{amsmath}

\usepackage{graphicx}
\usepackage[percent]{overpic}
\usepackage{multirow}
\usepackage{color}
\usepackage{rotating}

\newcommand{\msun}{${\rm M_{\sun}}$}

\def\ltsima{$\; \buildrel < \over \sim \;$}
\def\simlt{\lower.5ex\hbox{\ltsima}}
\def\gtsima{$\; \buildrel > \over \sim \;$}
\def\simgt{\lower.5ex\hbox{\gtsima}}
\def\kms{{\rm\,km\,s^{-1}}}
\def\pc{{\rm\,pc}}
\def\kpc{{\rm\,kpc}}

\def\msun{{\rm\,M_\odot}}

\newcommand{\fmmm}[1]{\mbox{$#1$}}

\newcommand{\mcnd}{\mbox{\fmmm{'}\hskip-0.3em .}}


\def\deg{^\circ}
\def\degg{\hbox{$\null^\circ$\hskip-3pt .}}
\def\sec{\hbox{"\hskip-3pt .}}

\def\Gyr{{\rm\,Gyr}}

\def\masyr{{\rm\,mas \, yr^{-1}}}

\def\ltsima{$\; \buildrel < \over \sim \;$}
\def\gtsima{$\; \buildrel > \over \sim \;$}

\def\Pal{Palomar~5}

\shorttitle{The Palomar 5 stellar stream}
\shortauthors{Ibata et al.}

\begin{document}

\title{Feeling the pull,\\ a study of natural Galactic accelerometers. II:\\
kinematics and mass of the delicate stellar stream of the Palomar 5 globular cluster\altaffilmark{1}}

\author{Rodrigo A. Ibata\altaffilmark{2}}
\author{Geraint F. Lewis\altaffilmark{3}}
\author{Guillaume Thomas\altaffilmark{2}}
\author{Nicolas F. Martin\altaffilmark{2,4}}
\author{Scott Chapman}

\altaffiltext{1}{Based on observations obtained with MegaPrime/MegaCam, a joint project of CFHT and CEA/DAPNIA, at the Canada-France-Hawaii Telescope (CFHT) which is operated by the National Research Council (NRC) of Canada, the Institute National des Sciences de l'Univers of the Centre National de la Recherche Scientifique of France, and the University of Hawaii. Based on observations made with ESO Telescopes at the La Silla Paranal Observatory under programmes ID 081.B-0258(A) and ID 083.B-0403(A)}

\altaffiltext{2}{Observatoire astronomique de Strasbourg, Universit\'e de Strasbourg, CNRS, UMR 7550, 11 rue de lÕUniversit\'e, F-67000 Strasbourg, France; rodrigo.ibata@astro.unistra.fr}
\altaffiltext{3}{Sydney Institute for Astronomy, School of Physics, A28, The University of Sydney, NSW, 2006, Australia}
\altaffiltext{4}{Max-Planck-Institut f\"ur Astronomie, K\"onigstuhl 17, D-69117 Heidelberg, Germany}

\begin{abstract}
We present two spectroscopic surveys of the tidal stellar stream of the Palomar 5 globular cluster, undertaken with the VLT/FLAMES and AAT/AAOmega instruments. We use these data in conjunction with photometric data presented in the previous contribution in this series to classify the survey stars in terms of their probability of belonging to the \Pal\ stellar stream. We find that high-probability candidates are only found in a very narrow spatial interval surrounding the locus of the stream on the sky. PanSTARRS RRLyrae stars in this region of sky are also distributed in a similar manner.  The absence of significant ``fanning'' of this stellar stream confirms that Palomar~5 does not follow a chaotic orbit. Previous studies have found that \Pal\ is largely devoid of low-mass stars, and we show that this is true also of the stellar populations along the trailing arm out to $6\deg$. Within this region, which contains 73\% of the detected stars, the population is statistically identical to the core, implying that the ejection of the low-mass stars occurred before the formation of the stream. We also present an updated structural model fit to the bound remnant, which yields a total mass of $4297\pm98 \msun$ and a tidal radius $0.145\pm0.009\kpc$. We estimate the mass of the observed system including the stream to be $12200\pm400\msun$, and the initial mass to have been $\sim47000\pm1500\msun$. These observational constraints will be employed in our next study to model the dynamics of the system in detail.
\end{abstract}


\section{Introduction}
\label{sec:Introduction}

Stellar streams from low-mass progenitors such as globular clusters provide us with a powerful tool to explore the mass distribution in galaxies. The path that a stream delineates on the sky probes directly the large-scale distribution of matter in the host, while small-scale inhomogeneities along the stream retain information about the distribution of substructure in the parent galaxy \citep{2002MNRAS.332..915I,2002ApJ...570..656J,2012ApJ...760...75C,2014ApJ...788..181N,2016PhRvL.116l1301B,2016MNRAS.460.2711T,2016arXiv160901282E,2016arXiv160901298B,2016MNRAS.463..102E}. The latter is of particular interest as it opens the possibility to constrain the preponderance of dark matter sub-haloes, and hence test one of the key predictions of $\Lambda$CDM theory.

In \citet[][hereafter Paper~I]{2016ApJ...819....1I}, we presented a comprehensive introduction to this problem, and argued that one of the most interesting systems to study in this context is the long and delicate stellar tidal stream of the Palomar~5 globular cluster. This cluster is found almost exactly above the Galactic bulge, half-way towards the North Galactic Pole, at $(\ell,b)=(0\degg85,+45\degg86)$, and at a Heliocentric distance of $23.6^{+0.8}_{-0.7}\kpc$ \citep{2015ApJ...803...80K}. The associated stellar stream emanates from the progenitor to form an immensely long structure, more than $10\kpc$ in extent, that is seen projected on the sky along the direction approximately parallel to the Galactic disk. The leading part of the stream extends towards negative $\ell$ (and to the West), where it disappears out of publicly accessible surveys. The trailing stream extends to positive $\ell$ (and towards the East), dropping below current detection levels $\sim 15\deg$ from the progenitor. The thinness and length of this stream make this one of the most interesting and promising cases to study at present.

In Paper I we presented two new photometric surveys, undertaken with the Canada France Hawaii Telescope (CFHT) and Kitt Peak National Observatory (KPNO) Mayall 4m telescope. These surveys probed the stream with good photometric precision over a larger field than has been achieved in previous studies. The CFHT g and r band data are particularly useful for detecting the numerous main sequence stars of the stream, while reducing the contamination from other Galactic stars as well as unresolved background galaxies. We found that there is no evidence for gaps in the stream up to a spatial scale of $2\deg$ $(\sim 1\kpc)$, a conclusion that should allow us to place useful limits on the number and properties of the $\Lambda$CDM sub-haloes that the stream has interacted with. Analysing the same dataset, \citet{2016arXiv160901282E} find evidence for two very large gaps ($\sim 2\deg$ and $\sim 9\deg$), which they claim are the scars of sub-halo flybys. We were also able to measure the relative distance of the stream population over the survey; a slight gradient was found such that in the outermost fields the trailing stream is $1.9\kpc$ more distant while the leading stream is $1.2\kpc$ closer to us. Interestingly, we measured the stream to be very thin in the most distant trailing stream fields. The stars in these regions should have been the first to be lost to the system, approximately $4\Gyr$ ago, so to have kept such a narrow structure ($58\pc$) over such a long period of time, suggests that the Galactic potential has been very smooth during the evolution of the stream. \citet{2015ApJ...799...28P} have shown that a stream similar to that of \Pal\ orbiting in the triaxial potential fitted by \citet{2010ApJ...714..229L} to the Sagittarius stream, would lead to a ``fanning'' of the stars towards the extremities of the stream (due to chaos); the fact that this is not observed in our data suggests that the Galactic halo cannot possess that particular shape.

\begin{figure*}
\begin{center}
\includegraphics[angle=0, viewport= 25 30 760 585, clip, width=\hsize]{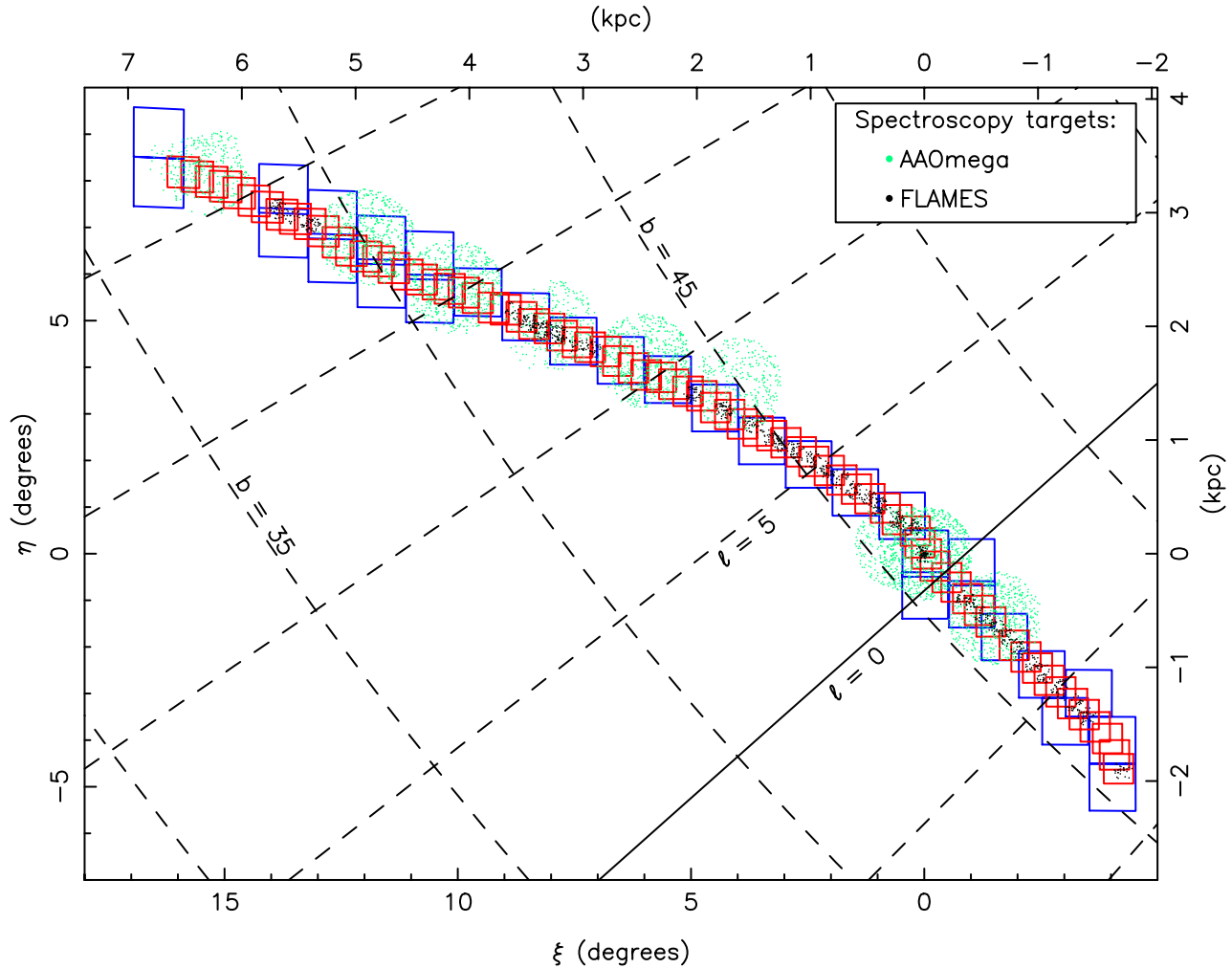}
\end{center}
\caption{Spectroscopic survey fields. The colored dots show the positions of the stars observed with AAOmega (green) and FLAMES (black). Most of these stars are encompassed within the photometric fields observed with the CFHT (blue) and KPNO (red). Additionally, the dashed lines mark a $5\deg\times5\deg$ grid in Galactic coordinates. Note that the $l=0\deg$ Galactic minor axis (which is shown with the continuous black line), passes very close to \Pal\, and is almost perpendicular to the stellar stream, implying that the structure is roughly parallel to the Galactic disk, albeit at a high Galactic latitude of $45\deg$. (This map is shown in standard gnomonic projection, with the $\xi$ and $\eta$ coordinates locally parallel to right ascension and declination, respectively).}
\label{fig:spectroscopic_fields}
\end{figure*}

In addition to the deep CFHT photometry, in Paper~I we also presented shallower KPNO photometry of this system using the M-band filter (approximately V) and the intermediate-band ``DDO~51'' filter (central wavelength 5145.2\AA, FWHM=162.9\AA) which covers the Mgb triplet. These data provide a means to constrain the surface gravity of the survey stars, allowing us to discriminate giants from contaminating dwarf stars.

The present contribution builds upon the survey of Paper~I. Here we will present a large spectroscopic survey of the stellar stream in Section~\ref{sec:Spectroscopic_Observations}, and explain in Section~\ref{sec:Decontamination} how the sample is decontaminated. The spatial distribution of the kinematically selected stars is examined in Section~\ref{sec:Spatial}. In Section~\ref{sec:Structure} we fit structural models to the remnant, estimate the mass in the bound versus unbound components as well as the initial mass of the system, and examine the differences in stellar populations along the stream. Finally in Section~\ref{sec:Conclusions} we draw the conclusions from this study.

\begin{table}
\begin{center}
\caption{Properties of the globular cluster \Pal.}
\label{tab:properties}
\begin{tabular}{ccc}
\tableline\tableline
Parameter & value & source \\
\tableline
RA & $15^h 16^m 05^s.3$ & 1\\
Dec & $-00\deg 06' 41\sec0$ & 1 \\
$\ell$ & $0.8522$ & \\
$b$   & $+45.8599$ & \\
$E({\rm B-V})$ & $0.06$ mag & 2 \\
$(m-M)_0$ & $16.86$ & 3 \\
Distance & $23.5\kpc$ &  \\
Angular scale & $411\pc$ per degree & \\
${\rm [Fe/H]}$ & -1.3 & 4 \\
Velocity $v_{helio}$ & $-58.7\pm0.2$ & 5 \\
Velocity dispersion $\sigma_{v}$ & $0.9\pm0.2$ & 5 \\
\tableline\tableline
\end{tabular}
\tablecomments{The sources are: 1 = \citet{DiCriscienzo:2006jv}, 2 = \citet{Schlegel:1998fw}, 3 = \citet{2011ApJ...738...74D}, 4 = \citet{2002AJ....123.1502S}, 5=\citet{2002AJ....124.1497O}. Rows without source information are derived from other table parameters.}
\end{center}
\end{table}

\section{Spectroscopic Observations}
\label{sec:Spectroscopic_Observations}

\subsection{FLAMES data}
\label{sec:FLAMES_data}

On 2009 Jun 23--27 we used the FLAMES multi-object spectrograph on the 8m VLT, to observe 35 fields along the Palomar 5 stellar stream. In ``Medusa'' mode it allows up to 130 targets to be allocated over a circular field of $12.5\arcmin$ radius, although typically $\sim 20$ fibers in each configuration were set aside for monitoring the spectrum of the sky. The high resolution setting HR21 was used, which straddles the \ion{Ca}{2} triplet feature and covers the spectral region between $8484\AA$ and $9001\AA$ with a resolution of $R=16200$. 

Each field consisted of $3\times600$~s exposures. The spectra were initially processed with the ``ESOREX'' package, which extracts and wavelength calibrates the spectra. After applying our own custom-made sky-subtraction algorithm, we measured the stellar radial velocities, velocity uncertainties and equivalent widths using a variant of the algorithm presented in \citet{2011ApJ...738..186I}. The quality of these data is $S/N\sim150$ at $g=15.5$, degrading to $S/N\sim6$ at $g=20$.

\begin{figure*}
\begin{center}
\includegraphics[angle=0, viewport= 50 55 780 565, clip, width=\hsize]{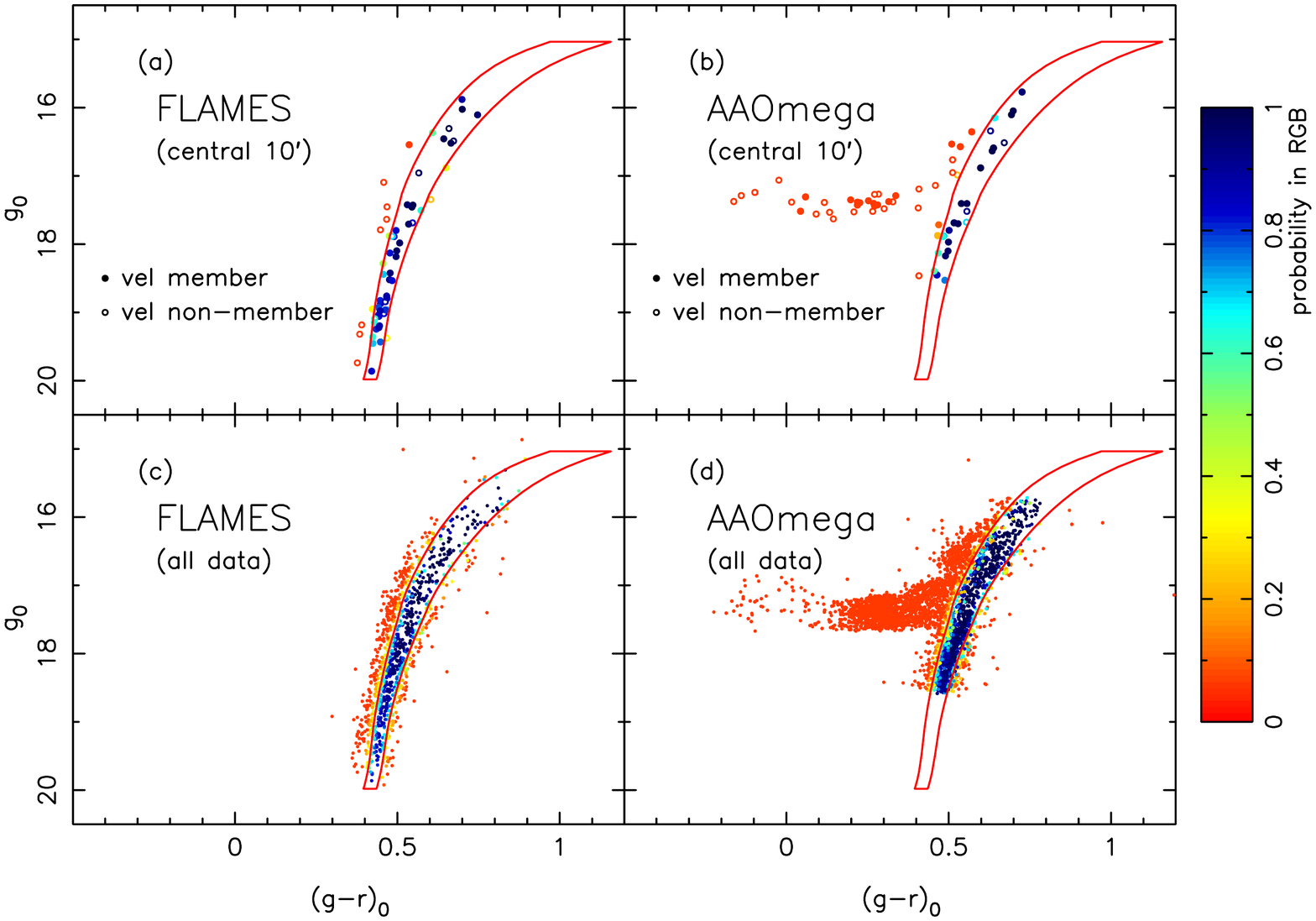}
\end{center}
\caption{The RGB selection adopted. The red polygon in each panel shows a broadened Dartmouth isochrone model with ${\rm [Fe/H]=-1.8}$, ${\rm [\alpha/Fe]=+0.6}$ and age $11.5\Gyr$; we stress that this model simply provides a convenient visual fit to the RGB, and we are not using it to make any claims about the physical properties of the cluster. We chose the width of the RGB to be a quadratic as a function of magnitude, to encompass the obvious RGB seen in (a). That panel displays the targeted stars in the FLAMES survey that are located near the center of the cluster remnant (within $10\arcmin$), where the contamination is least problematic. The filled circles show stars that have velocities within $10\kms$ of the cluster, which we consider to be velocity members, while the open circles have velocities beyond that range. The chosen RGB polygon can be seen to encompass the majority of the member stars. The color coding assigned to the points corresponds to a probability of being within the RGB region, assuming $0.01$ magnitude errors in ${\rm g}$ and ${\rm r}$ (these are taken to approximate the photometric zero-point errors between fields, as well as allowing for some uncertainty in the RGB model). Panel (b) shows the same information for the AAOmega survey. Note that we included many blue horizontal branch candidate stars among the AAOmega targets. Panels (c) and (d) show the same information as the upper panels, but for the entire survey region. In calculating the RGB membership probability, we have taken into account the distance gradient as a function of position along the stream (which translates into small magnitude shifts).}
\label{fig:RGB_selection}
\end{figure*}

The FLAMES fields are fully contained within the CFHT survey region (see Figure~\ref{fig:spectroscopic_fields}), and yielded a total of 1327 stars.

\subsection{AAOmega data}
\label{sec:AAOmega_data}

We also observed 15 fields with the AAOmega multi-object spectrograph at the 4m Australian Astronomical Telescope on the nights of 2006 June 13--18. This instrument uses a robot positioner to allocate its 400 fibers over a circular field of radius $1\deg$. The locations of the 15 fields (which have 9 different central positions) are shown in Figure~\ref{fig:spectroscopic_fields}. In each configuration several tens of fibers were assigned to blank regions of sky so as to allow for accurate sky subtraction.

The 1700D grating was used to measure the spectral region between $\sim 8400$\AA\ and $\sim 8850$\AA, at a resolution of $0.24$\AA$\, {\rm pixel^{-1}}$. In each field, three exposures of $1800$~s were combined, providing $S/N\sim60$ (per pixel) at $g=16$, deteriorating to $S/N\sim5$--$10$ for the faintest targets at $g=18.8$.

The spectra were extracted, wavelength calibrated and sky-subtracted using the ``2dfdr'' software\footnote{\tt https://www.aao.gov.au/science/software/2dfdr}. Subsequently, the velocities and equivalent widths of the stars were measured with the same custom-made software as was used for the FLAMES spectra. The resulting catalogue contains 4567 stars, with 3642 having velocity errors $<5\kms$ and Heliocentric radial velocity $|v_{helio}| < 300\kms$. 
Selecting good quality stars from both datasets ($S/N>40$ for FLAMES) ($S/N>30$ for AAOmega), we obtain 45 stars in common. The velocity difference between the two samples is $v_{\rm FLAMES}-v_{\rm AAO} = 3.73 \pm 1.89\kms$, and we used this mean offset to put the AAO velocities onto the FLAMES zero-point.

\begin{figure*}
\begin{center}
\hbox{
\includegraphics[angle=0, viewport= 45 55 770 585, clip, width=\hsize]{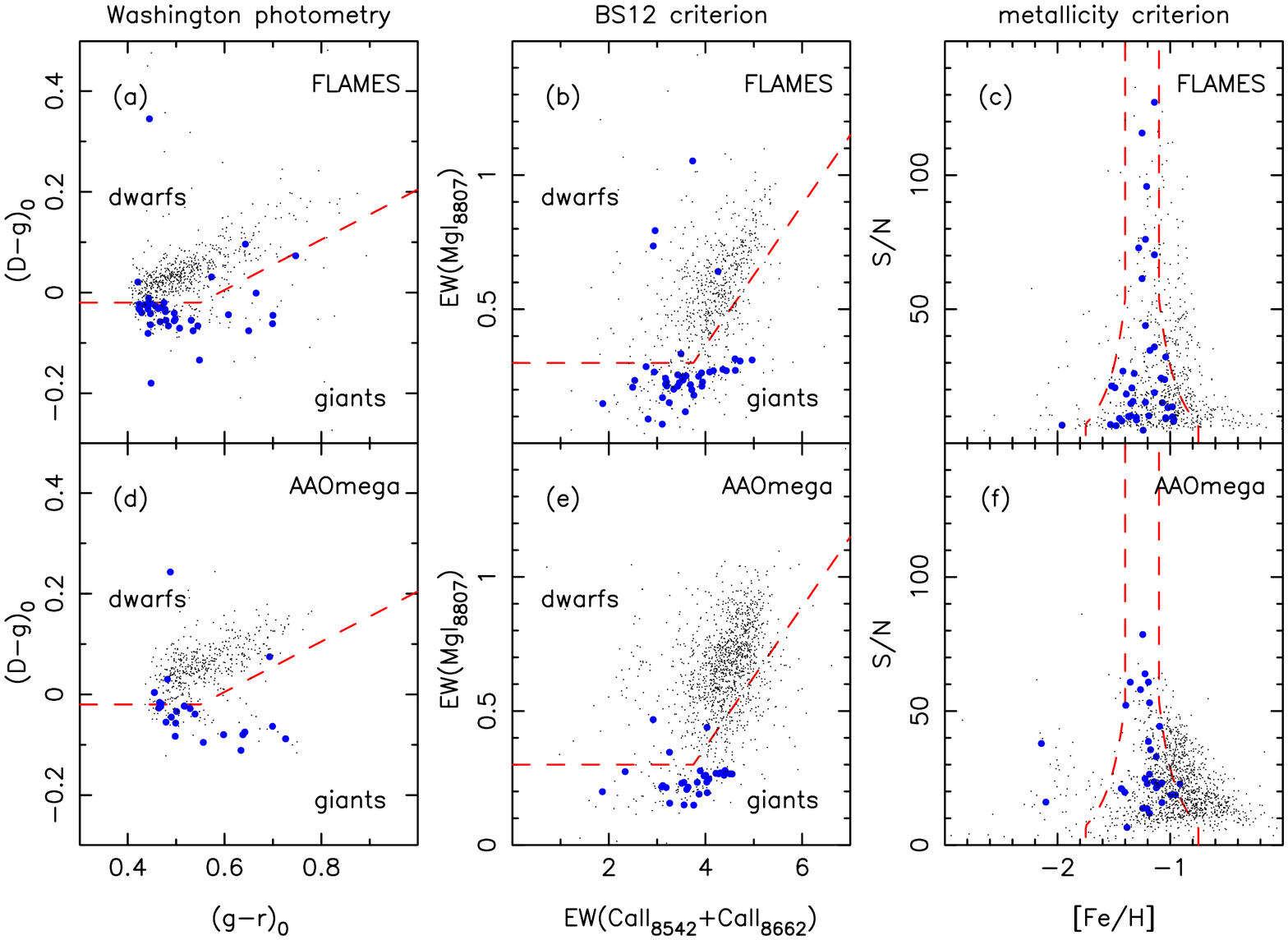}
}
\end{center}
\caption{Correlations between stellar properties. All stars have been selected to be plausible RGB members (probability $P_{\rm RGB}>0.1$), and to have velocity uncertainties smaller than $5\kms$. The blue circles in all panels represent stars within $10\arcmin$ of the cluster center and which are also radial velocity members ($|v-v_{\rm Pal5}| < 10\kms$). In panels (a) and (d) we show the giant-dwarf discrimination based on the DDO~51 photometry. The dashed line shows our visually selected boundary to separate giants (below the line) from dwarfs (above the line). Panels (b) and (e) likewise show the giant-dwarf discrimination, but this time using the line index criteria of \citet{2012A&A...539A.123B} (together with their boundary). As before, the dwarfs lie above the boundary. Finally, in (c) and (f), we display our selection based on our spectroscopic metallicity measurements.}
\label{fig:correlations}
\end{figure*}

\section{Decontamination and Velocity Fidelity}
\label{sec:Decontamination}

Despite having access to this very large sample of $\sim 5000$ stars, our initial analysis of the dataset showed that the number of stream stars that could be identified easily was very small, so we were obliged to develop a method to de-contaminate the sample, which we will describe in the present section.

The first criterion we use is to limit the sample according to their proper motion, as given by the SDSS DR12 ``Stars'' catalog \citep{Alam:2015go}. For this, we reject all stars that have proper motions $|\mu-\mu_{\rm Pal 5}| > 5\masyr + 2 \delta \mu$, where $\mu$ is the measured proper motion of an SDSS star, $\delta \mu$ is the corresponding uncertainty, and $\mu_{\rm Pal 5}$ is the proper motion of \Pal\ (taken to be $\mu_\alpha=-2.296\pm0.186\masyr$, $\mu_\delta=-2.257\pm0.181\masyr$; \citealt{2015ApJ...811..123F}). The $5\masyr$ limit (i.e. $>500\kms$ at the distance of \Pal) encompasses all plausible variations in velocity of the stream, and allows us to reject $\sim 30$\% of the targeted stars. However, this quality cut turned out to be relatively unimportant to the present analysis, as it does not provide additional discrimination beyond what we obtain from the RGB star selection detailed below.

\subsection{RGB CMD selection}

The initial targets were selected in a rather broad box around the cluster fiducial (the selection was made before the analysis of Paper~I, and we wanted to make sure that our sample was not biassed by possible variations in stream distance). However, since the contamination is our primary challenge, we can now obtain a cleaner sample of RGB stars by choosing stars within a narrower box that follows the measured distance gradient. The adopted RGB selection region is shown in panels (a) and (b) of Figure~\ref{fig:RGB_selection}, where we also show the CMD of the central region of 10\arcmin\ radius around the globular cluster. The box is a broadened Dartmouth stellar population model \citep{2008ApJS..178...89D} of age $11.5\Gyr$, metallicity ${\rm [Fe/H]=-1.8}$ and alpha abundance ${\rm [\alpha/Fe]=+0.6}$, which encompasses almost all the stars that have velocities within $10\kms$ of the cluster mean ($-58.7\pm0.2\kms$; \citealt{2003AJ....126.2385O}). Note that the adopted Dartmouth model was not rigorously fitted to the data, but rather it was chosen as it gave the most satisfactory visual approximation to the central sample. In particular, the metallicity of this model is much lower than spectroscopic studies indicate \citep{2002AJ....123.1502S}. 

The data in Figure~\ref{fig:RGB_selection} are colored according to their probability of falling within the RGB selection box. In the upper two panels, this is simply determined from the photometric uncertainty (assuming Gaussian photometric uncertainties and a minimum error of 0.01 mags in the g and r bands to reflect the zero-point uncertainties). The bottom two panels display the data from the full survey, and the RGB probability also includes the slight magnitude shift, as a function of angular distance away from the cluster, to account for the line of sight distance measurements presented in Paper~I.

\subsection{Other membership criteria}

In addition to the ${\rm (g-r,g)}$ CMD selection for RGB members, we also implement a photometric selection based on Washington photometry, a selection on the equivalent width of the \ion{Mg}{1} 8807\AA\ line, and a selection based on spectroscopic metallicity. The workings of these three additional tests are summarized in Figure~\ref{fig:correlations}. Panels (a) and (d) show the Washington photometry selections applied to the FLAMES and AAOmega data, respectively. This ${\rm g-r}$ vs ${\rm D-g}$  color-color diagram is an attempt to recreate similar diagnostic diagrams used in earlier studies (e.g. \citealt{2000AJ....120.2550M}). As explained in Paper~I, we were not able to calibrate the zero-point of our DDO~51 band (which we label ``D''), so we are forced to make our own empirical cuts in this color-color diagram to separate dwarfs and giants. The large blue dots in the panels show the parameter values of RGB members of the cluster (with RGB probability $P_{\rm RGB}>0.1$ according to the criterion of Figure~\ref{fig:RGB_selection}) and that also have velocities within $10\kms$ of the cluster mean. These giants lie preferentially below the dashed red line, which we use as our empirical dwarf/giant boundary. For each star, using its corresponding photometric uncertainties, we calculate the probability $P_{\rm DDO 51}$ that it lies within the parameter region inhabited by giants.

We also employ the dwarf/giant discrimination method developed by \citep[hereafter BS12]{2012A&A...539A.123B} that uses the gravity sensitive \ion{Mg}{1} 8807\AA\ line. This is shown in panels (b) and (e), along with the boundary line defined by BS12. Again, the the stars in the survey are given a probability $P_{\rm BS12}$ of being a giant according to their position in this diagram (and accounting for the equivalent width measurement uncertainties).

Finally, in panels (c) and (f) we present the metallicity measurements for the FLAMES and AAOmega data, respectively, based on the \ion{Ca}{2} equivalent widths. From our measurements, the velocity members of the cluster (blue dots) have mean ${\rm [Fe/H]=-1.25}$. We define the metallicity selection region within the red dashed lines ($\pm 0.15 \, {\rm dex}$ around the mean value), which we widen as shown for spectra of low signal to noise.

\begin{figure}
\begin{center}
\includegraphics[angle=0, viewport= 10 5 410 395, clip, width=\hsize]{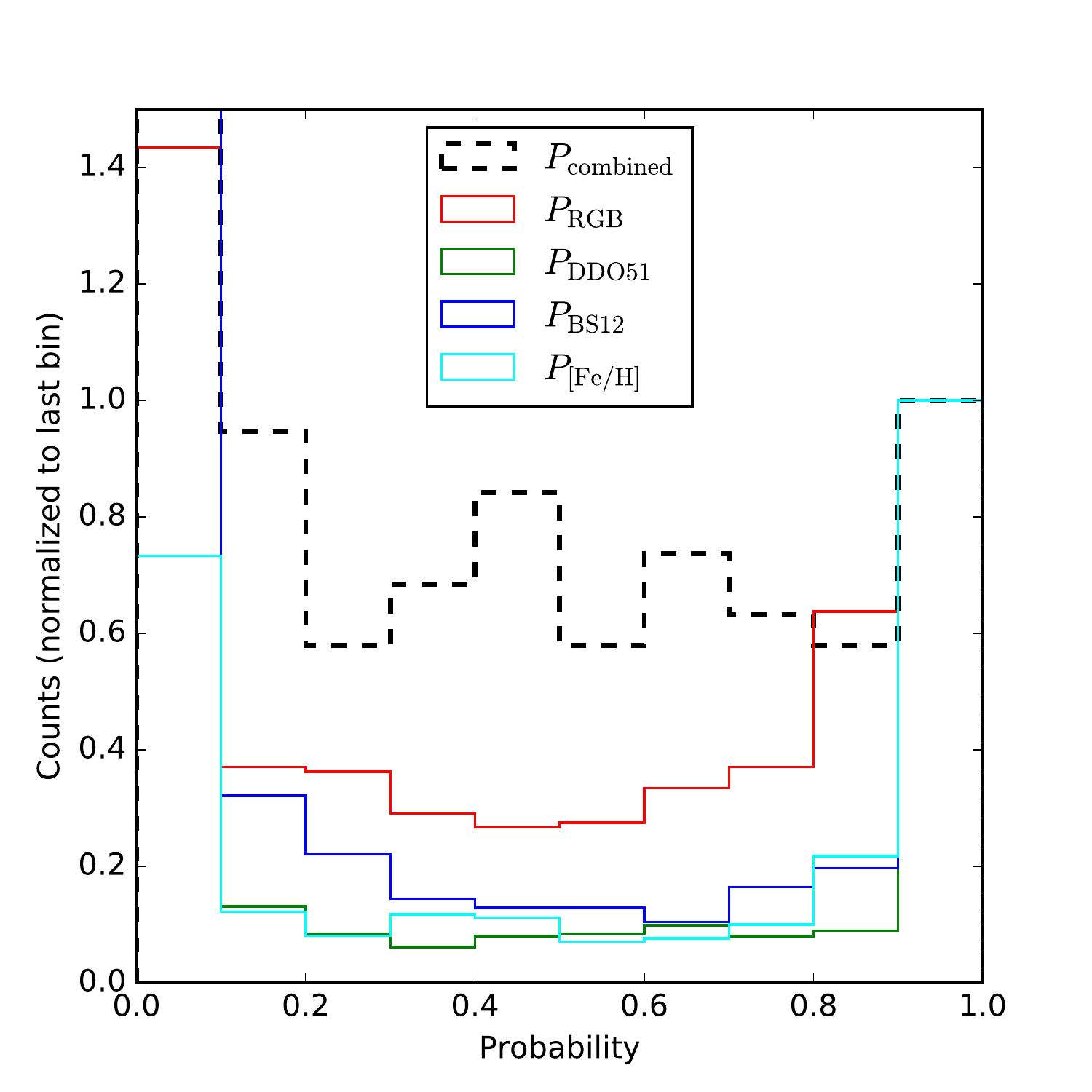}
\end{center}
\caption{Distribution of probability values for the four membership criteria, as derived for the FLAMES sample. The histograms have been normalized to the value in the last bin. The dashed line histogram is the distribution of the product of the four probabilities (the first bin of the $P_{\rm combined}$ histogram has a value of 64.0 with this normalization, far beyond the plotted limit).}
\label{fig:prob}
\end{figure}

Each of the four membership criteria (RGB position, Washington photometry color, BS12 criterion, and metallicity) allow us to calculate a probability according to how far a star's parameter measurements lie from the defined criterion boundary. In Figure~\ref{fig:prob} we display the distribution of these probability values. The four criteria can be seen to possess two peaks: one near zero and another peak at probability values $\simgt 0.8$. The product of the four probabilities is shown with the dashed-line histogram. In the combined probability histogram, the contrast of the high-probability peak is much diminished. This is not surprising however, since say a star that has a membership probability of 80\% according to each of the individual criteria, would have $P_{\rm combined} = 0.7^4 \sim 0.4$. We tentatively choose those stars with $P_{\rm combined}>0.2$ as possible stream members, since this appears to be the location of a break between them and the clear non-members (which are highly peaked near zero).

\subsection{Comparisons to earlier data}

We next attempt to ascertain the reliability of the velocities and dwarf/giant discrimination determined here by comparing our results to those presented in previous studies.

A study with particularly accurate velocities was that made by \citet{Odenkirchen:2009js} with the VLT high resolution spectrograph UVES. There are 12 stars in common between our FLAMES sample and that of  \citet{Odenkirchen:2009js}; these give a satisfactory offset of $v_{\rm FLAMES}-v_{\rm Odenkirchen} = 1.63 \pm 3.05\kms$. The stars in the \citet{Odenkirchen:2009js} study were classified as giants from an inspection of the width of the Mg `b' triplet feature in their high-resolution spectra. This gives us an opportunity to also check the reliability of our dwarf/giant discrimination: we find that all 12 stars pass the Washington photometry criterion ($P_{\rm DDO 51}>0.75$), while 11 pass the \citet{2012A&A...539A.123B} test (having $P_{\rm BS12}>0.90$). The star that does not pass the BS12 test (object 30017 in the \citealt{Odenkirchen:2009js} sample) has $P_{\rm BS12}=0$, and interestingly is a velocity non-member, which lends additional credence to our classification procedure.

Between our AAOmega sample and that of \citet{Odenkirchen:2009js} we find 9 stars in common. After rejection of one outlier we derive an offset of $v_{\rm AAO}-v_{\rm Odenkirchen} = 3.45 \pm 4.28\kms$.

Between the \citet{2015MNRAS.446.3297K} study and our AAOmega sample, there are 62 objects in common, of which 4 have very discrepant velocities $>10\kms$. One of these is a very blue star with $(g-r)_0=0.044$, and hence could be a radial velocity variable on the horizontal branch. We carefully examined the spectra of the other 3 stars, and found no reason to suspect anything wrong with the observations or our analysis, and the fits to the \ion{Ca}{2} lines performed by our software were clearly reasonable. Ignoring these 4 discrepant stars gives good consistency between the two surveys: $v_{\rm AAO}-v_{\rm Kuzma} = 2.00 \pm 2.53\kms$.

Note that there are no stars in the \citet{2015MNRAS.446.3297K} sample that are not in our AAOmega sample, apart from 3 of their listed stars, which happen to also be in the \citet{Odenkirchen:2009js} sample. Hence the \citet{2015MNRAS.446.3297K} sample provides no additional independent information over what we present here.

\section{Spatial and kinematic distribution}
\label{sec:Spatial}

\begin{figure}
\begin{center}
\hbox{
\vbox{
\includegraphics[angle=0, viewport= 200 170 775 580, clip, width=\hsize]{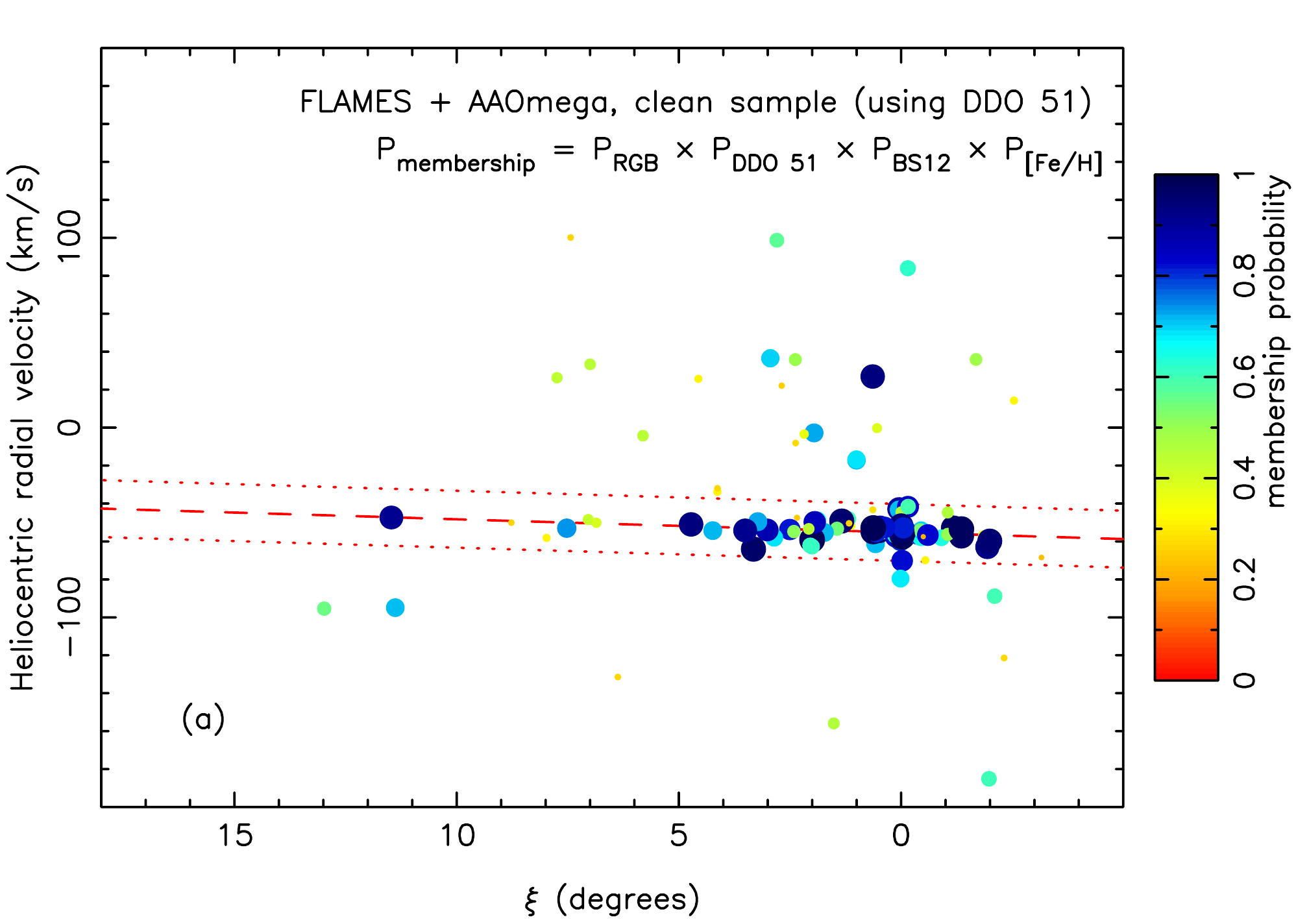}
\includegraphics[angle=0, viewport= 200 115 775 560, clip, width=\hsize]{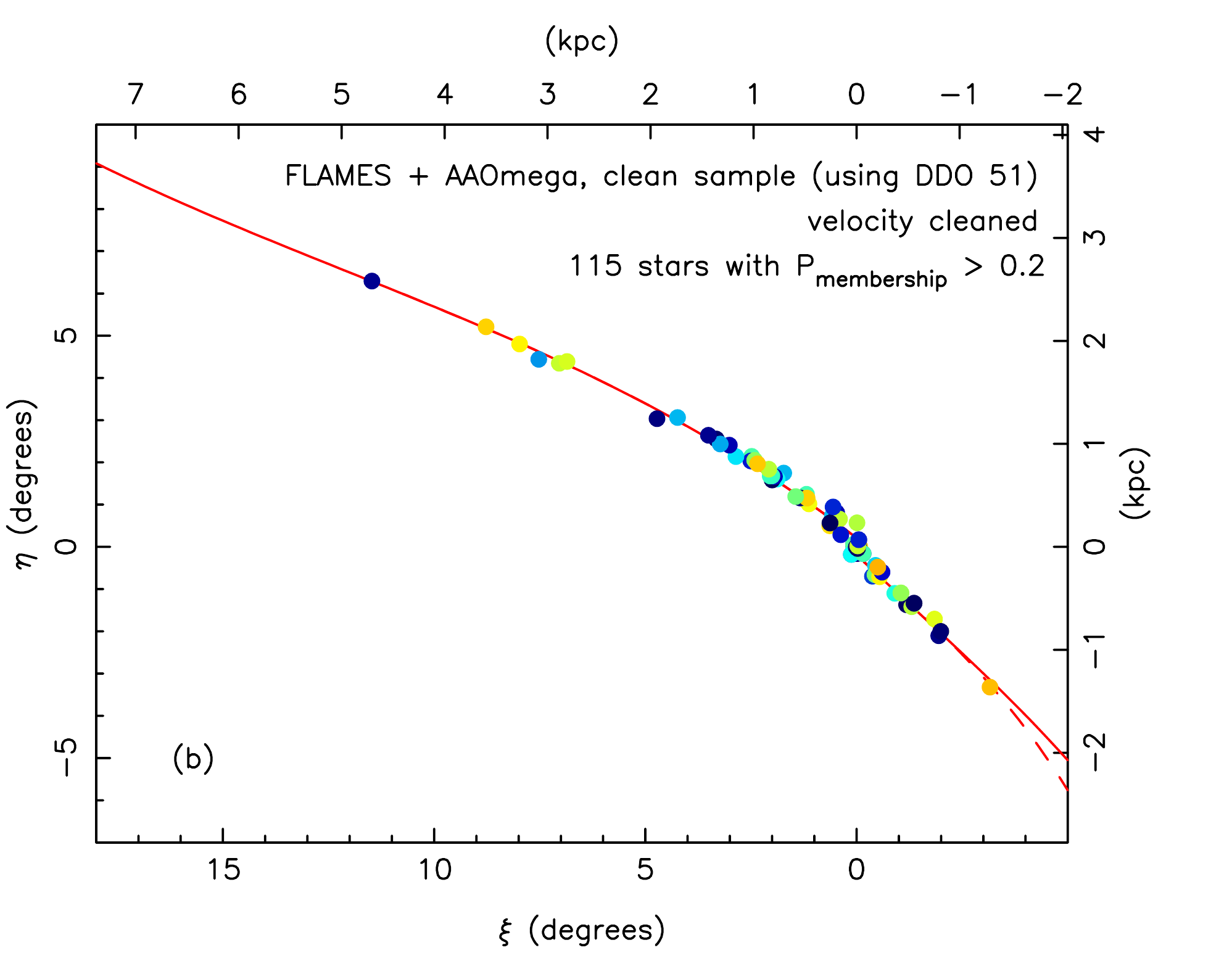}
}}
\end{center}
\caption{(a) : Heliocentric velocity as a function of standard coordinate $\xi$.  All stars have been selected to be plausible \Pal\ members (probability $P_{\rm membership}>0.2$), and to have velocity uncertainties better than $5\kms$. The membership probability of each star is shown in the color code as well as by the size of the circle, and is calculated from the product of the probabilities of the star belonging to the RGB, being a giant according to the Washington photometry criterion, being a giant according to the \citet{2012A&A...539A.123B} criterion, and being a member according to its metallicity. We have fitted a straight-line model to these data, using the ``conservative formulation'' of \citet{Sivia:2006us} to automatically reject outliers; the corresponding fit is shown with a dashed line, while the dotted lines show an interval of $\pm 15\kms$ around this fit. (b) : Sky positions of probable \Pal\ member stars. The sample is the same as that of panel (a), but is now curtailed to the region within $\pm 15\kms$ of the fitted gradient (between the dotted lines in that diagram). For comparison, in red, we have superimposed the cubic fit to the matched filter map presented in Paper~I. These high-likelihood spectroscopic members follow very closely the track of the photometric stream. The color coding of the points is identical to that of panel (a), marking membership probability.}
\label{fig:sel_A}
\end{figure}

In Figure~\ref{fig:sel_A} we show the main observational result of this survey: the 3-dimensional phase-space position of the most likely RGB stars. Panels (a) and (b) show, as a function of standard coordinate $\xi$, the radial velocity and the standard coordinate $\eta$ position, respectively. In both cases the data are color-coded according to the membership probability. Here we present the cleanest sample, using all of the RGB membership criteria, with $P_{\rm membership}=P_{\rm RGB} \times P_{\rm DDO 51} \times P_{\rm BS12} \times P_{\rm [Fe/H]} > 0.2$. This sample consists of 154 stars (which are listed in Table~\ref{tab:kinematics}), of which 116 have velocities in the range $-80\kms < v < -40\kms$; 72 of these stars were not previously known to be kinematic stream members.

\begin{figure}
\begin{center}
\hbox{
\vbox{
\includegraphics[angle=0, viewport= 200 170 775 580, clip, width=\hsize]{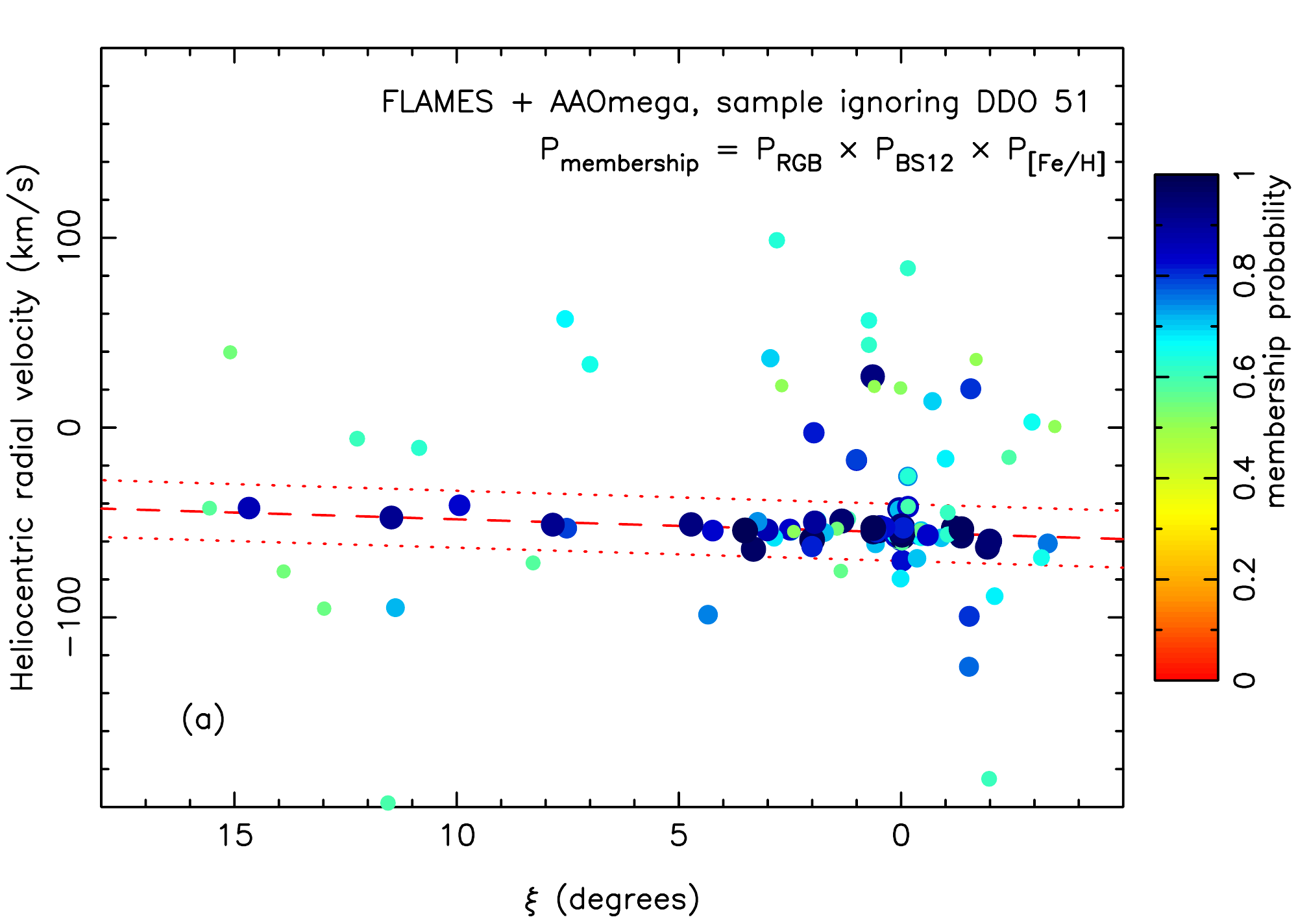}
\includegraphics[angle=0, viewport= 200 115 775 560, clip, width=\hsize]{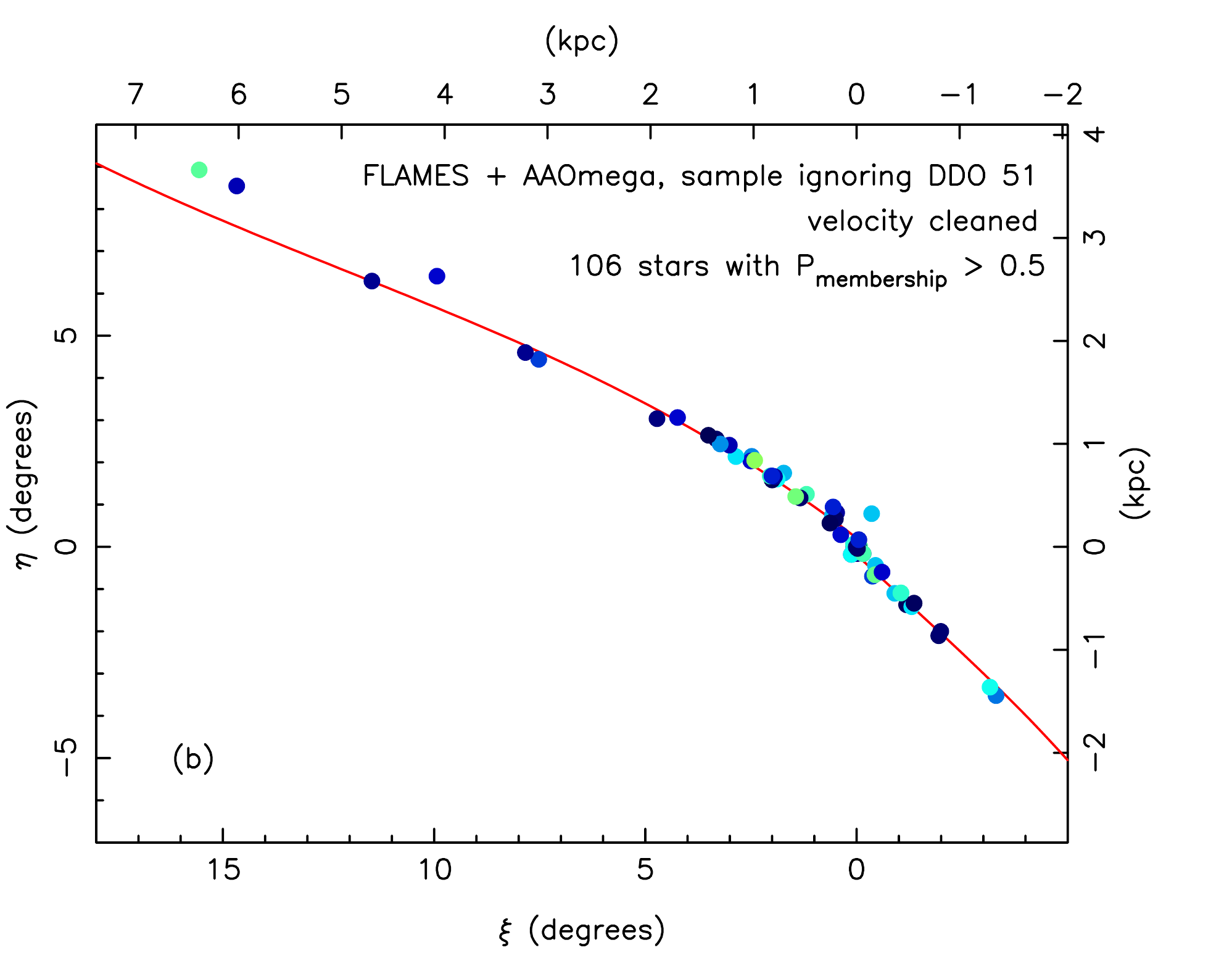}
}}
\end{center}
\caption{As Figure~\ref{fig:sel_A}, but ignoring the Washington photometry information. This alternative selection is presented in order to analyse the complete AAOmega dataset, as our DDO~51 survey does not fully cover all of the AAOmega fields (see Figure~\ref{fig:spectroscopic_fields}). While this should result in a lower fidelity sample than that of Figure~\ref{fig:sel_A}, it allows us to examine better the possible spread perpendicular to the stream. Note that we use $P_{\rm membership}>0.5$ for the present selection. To aid visual comparison, the lines in panels (a) and (b) are copied from Figure~\ref{fig:sel_A}.}
\label{fig:sel_B}
\end{figure}

The red dashed line in (a) shows a straight-line fit to this dataset, using the Bayesian automatic outlier rejection algorithm of \citet{Sivia:2006us} (their ``conservative formulation''). This fit is centered on $-55.30\kms$ and has a slope of $0.699 \kms \, {\rm degree^{-1}}$. The straight red dashed line can be seen to give a close representation of this dataset, out to $\xi \sim5\deg$ from the cluster, and even beyond $5\deg$ several stars line up very well along it out to $\xi =11\degg5$. The dotted lines mark the range of $\pm 15\kms$ of this fit, and the sky positions of all stars encompassed between these lines are shown in panel (b). The correspondence between these positions and the cubic polynomial stream fit presented in Paper~I (dashed red line) is striking.

\citet{2016MNRAS.463.1759B} recently presented a matched-filter map from the PanSTARRS survey, which showed that the leading arm of the \Pal\ stream continues on $\sim 1\degg5$ beyond the end of our CFHT survey, to $\xi=-6\deg$. From a FITS image of the PanSTARRS matched filter map (kindly provided by E. Bernard), at $\xi=-6\deg$ we measure $\eta=-6\degg25\pm0\degg1$. Adding this datum into our polynomial fit of the leading arm stream from Paper~I, we obtain a slightly modified track:
\begin{equation}
\eta_{\rm leading}(\xi) =-0.184 + 0.957 \xi + 0.0217 \xi^2 + 0.00502 \xi^3 \, .
\end{equation}
This improved model is shown as the solid red line in Figure~\ref{fig:sel_A}b, and will be used henceforth.

\begin{figure}
\begin{center}
\hbox{
\vbox{
\includegraphics[angle=0, viewport= 200 170 775 580, clip, width=\hsize]{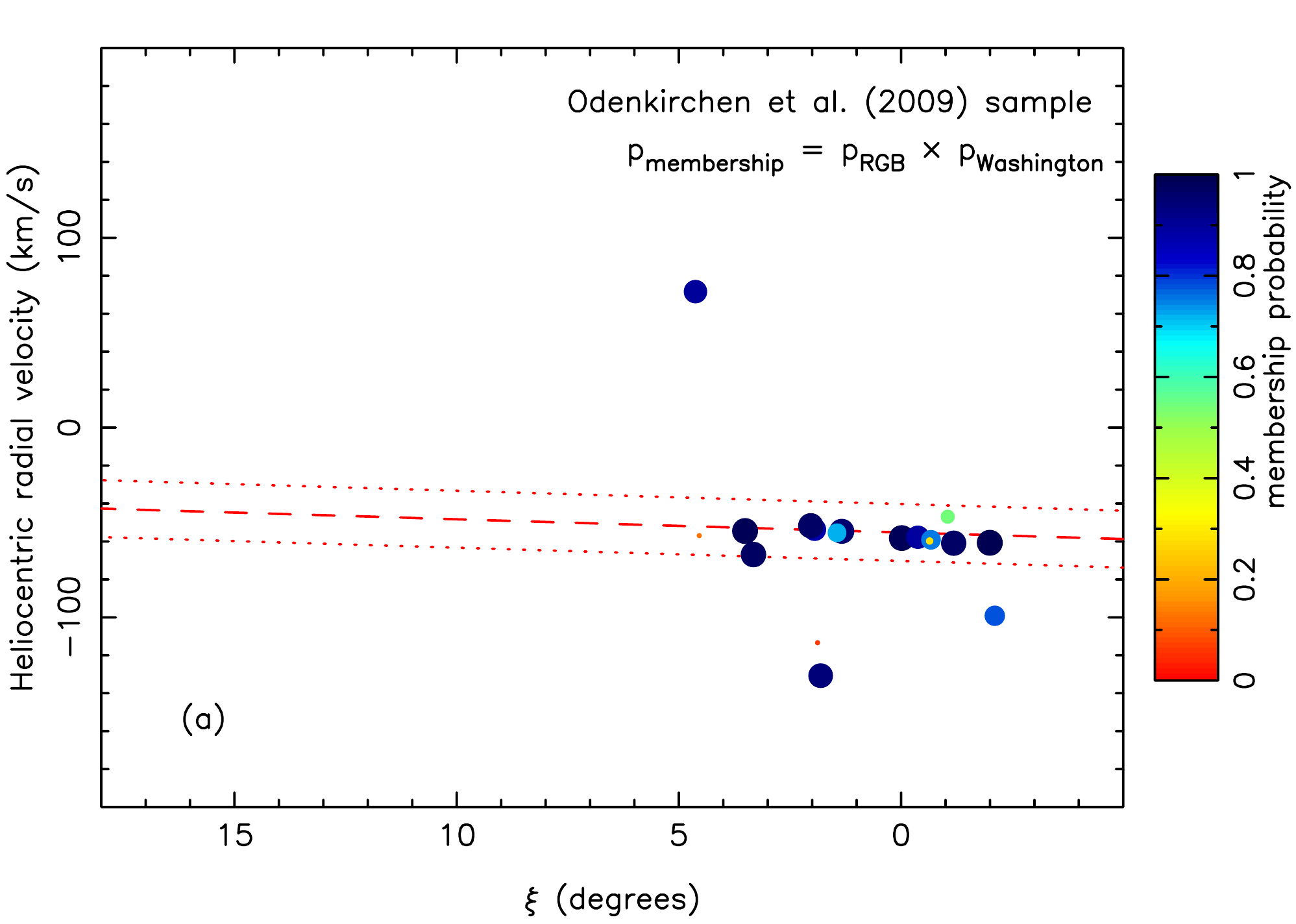}
\includegraphics[angle=0, viewport= 200 115 775 560, clip, width=\hsize]{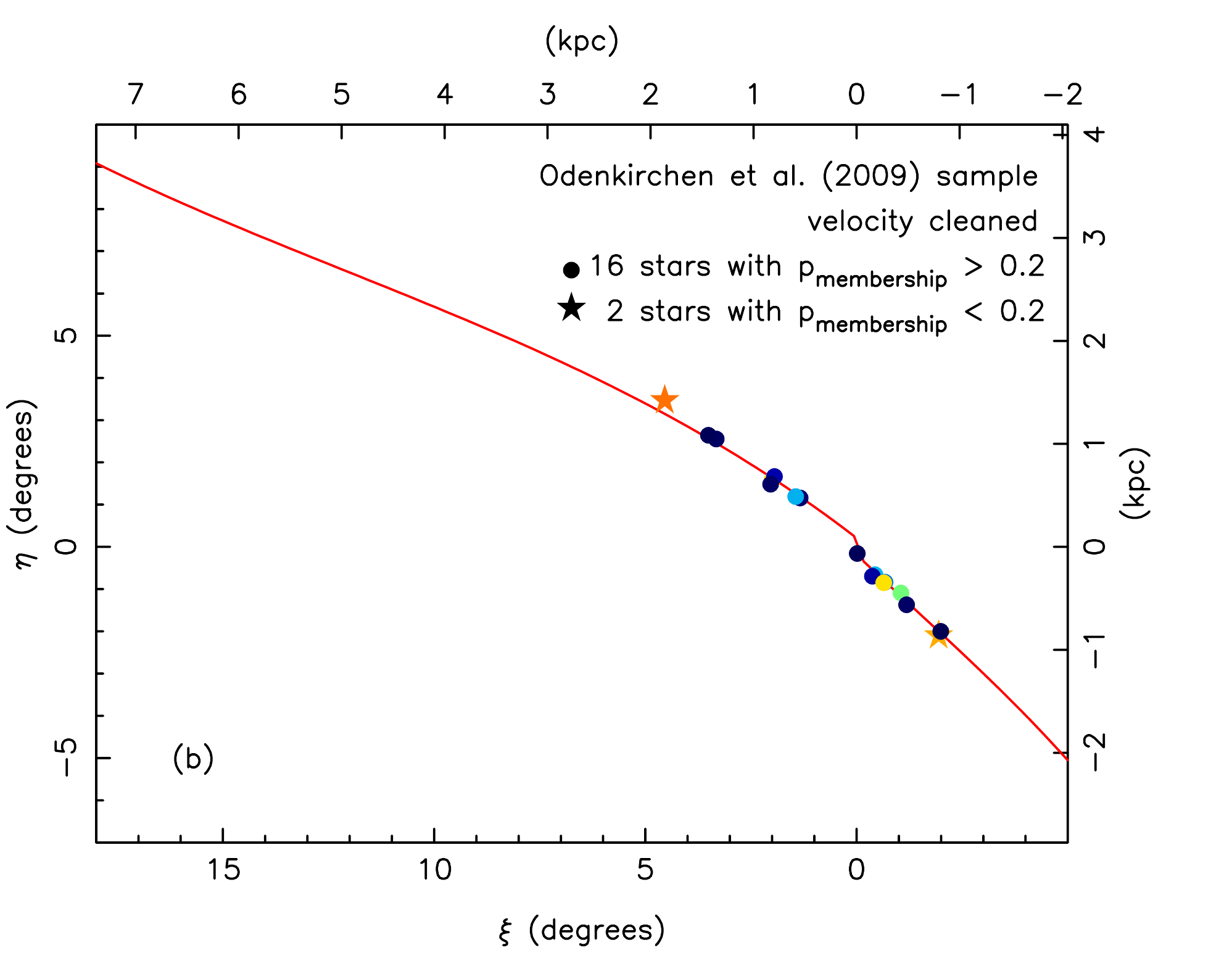}
}}
\end{center}
\caption{As Figure~\ref{fig:sel_A}, but for the 22 stars of the \citet{Odenkirchen:2009js} sample. The membership probability criterion used here is $P_{\rm RGB} \times P_{\rm DDO 51}$, since we do not have access to the relevant spectral information to calculate $P_{\rm BS12}$ or $P_{\rm [Fe/H]}$ for all the sample. Panel (b) shows the 18 stars within the velocity selection box of Figure~\ref{fig:sel_A}a. The positions of two likely non-member stars (with $P_{\rm membership}<0.2)$ are shown with pentagonal ``star'' glyphs. One can see that the high-likelihood stars follow very closely the fit to the matched filter map (red line).}
\label{fig:sel_Oden}
\end{figure}

We caution the reader that all the stars in the sample shown in Figure~\ref{fig:sel_A} are required to have DDO~51 photometry (otherwise $P_{\rm membership}=0$, since $P_{\rm DDO 51}=0$), and the footprint of that survey does not stray far from the solid red line fit. To overcome this limitation and probe the distribution of sources beyond the higher density stream region, we must abandon the Washington photometry criterion. The resulting sample is shown in Figure~\ref{fig:sel_B}, where we take $P_{\rm membership}=P_{\rm RGB} \times P_{\rm BS12} \times P_{\rm [Fe/H]} > 0.5$. A higher limiting probability is adopted since we are now dropping one of the probability factors. The straight-line fit from Figure~\ref{fig:sel_A}a is reproduced here. It is interesting to note that several new stars lie on the extrapolation of that line. The sky positions of the targets within $\pm15\kms$ of the fit are shown in Figure~\ref{fig:sel_B}b, and tantalisingly, the few additional objects that appear at large distance from the cluster center appear to be offset from the fiducial stream path (i.e. the solid red line).

\begin{figure}
\begin{center}
\hbox{
\vbox{
\includegraphics[angle=0, viewport= 200 170 775 580, clip, width=\hsize]{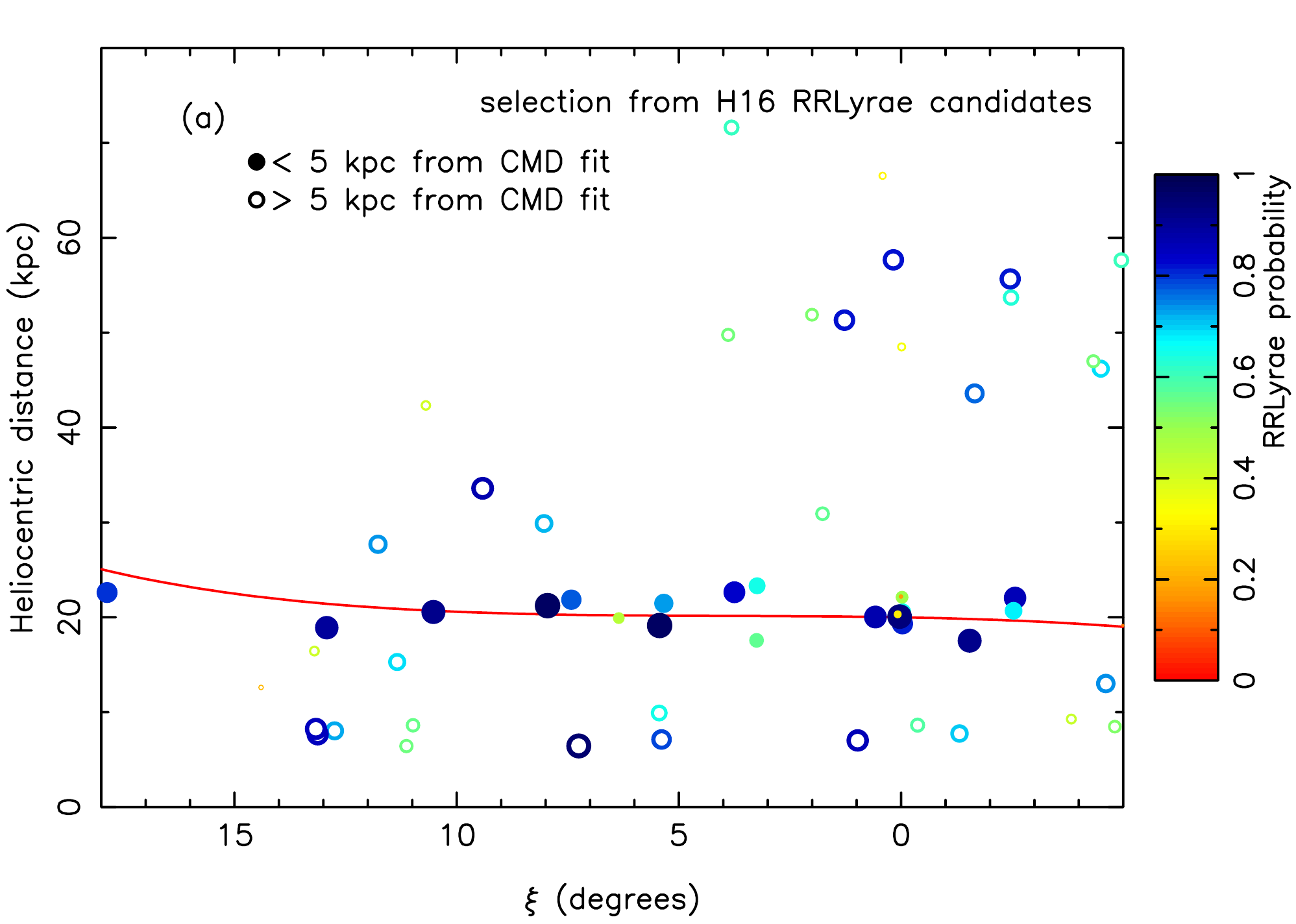}
\includegraphics[angle=0, viewport= 200 115 775 560, clip, width=\hsize]{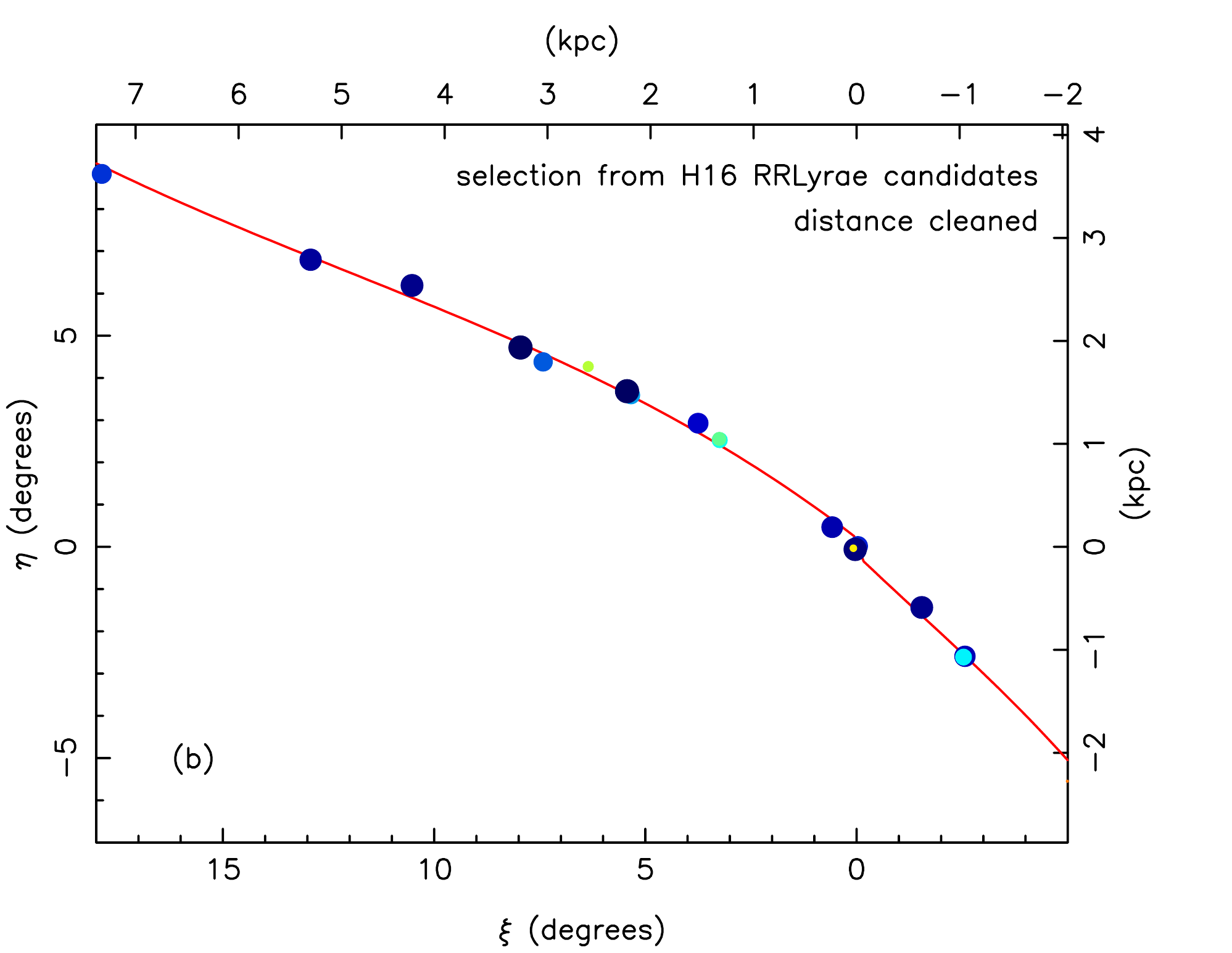}
}}
\end{center}
\caption{Distance and sky distribution of RRLyrae candidate stars. We show here all objects from the H16 PanSTARRS analysis with plausible RRLyrae-like probabilities ($P_{\rm RRLyrae}>0.1$),  that lie within $<0\degg3$ from the fitted stream path. In both panels the color and size of the points encodes the probability $P_{\rm RRLyrae}$. In (a) one can clearly perceive a good correspondence between the most probable RRLyrae candidates and the relative distance profile measured in Paper~I (red line), where the conversion to absolute distance adopts the Palomar~5 distance of $23.5\kpc$ fit by \citealt{2011ApJ...738...74D}. In (b) we show the spatial distribution of the stars within $5\kpc$ of the distance fit from Paper~I.}
\label{fig:RRLyrae1}
\end{figure}

\begin{figure}
\begin{center}
\hbox{
\vbox{
\includegraphics[angle=0, viewport= 200 170 775 580, clip, width=\hsize]{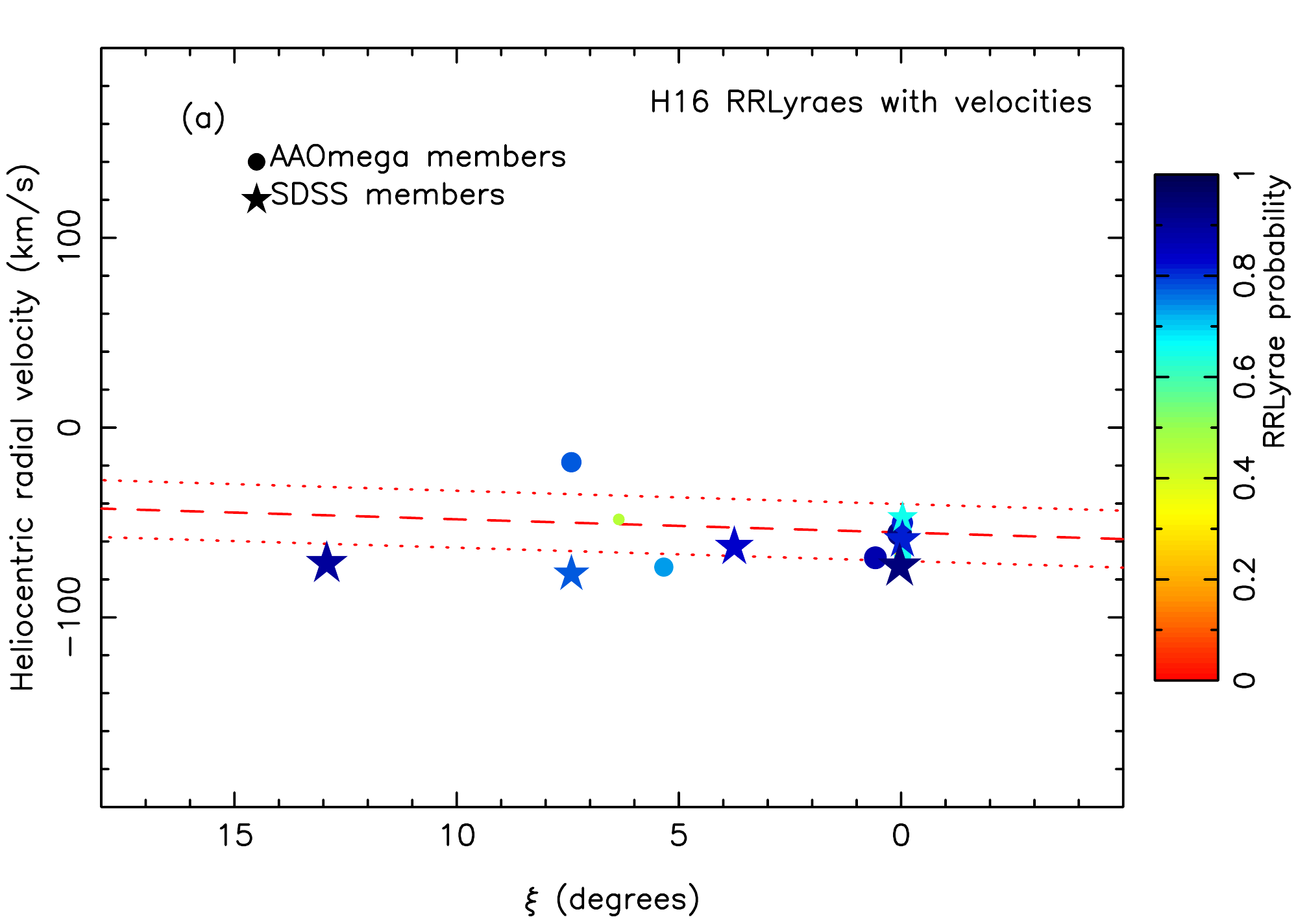}
\includegraphics[angle=0, viewport= 200 115 775 560, clip, width=\hsize]{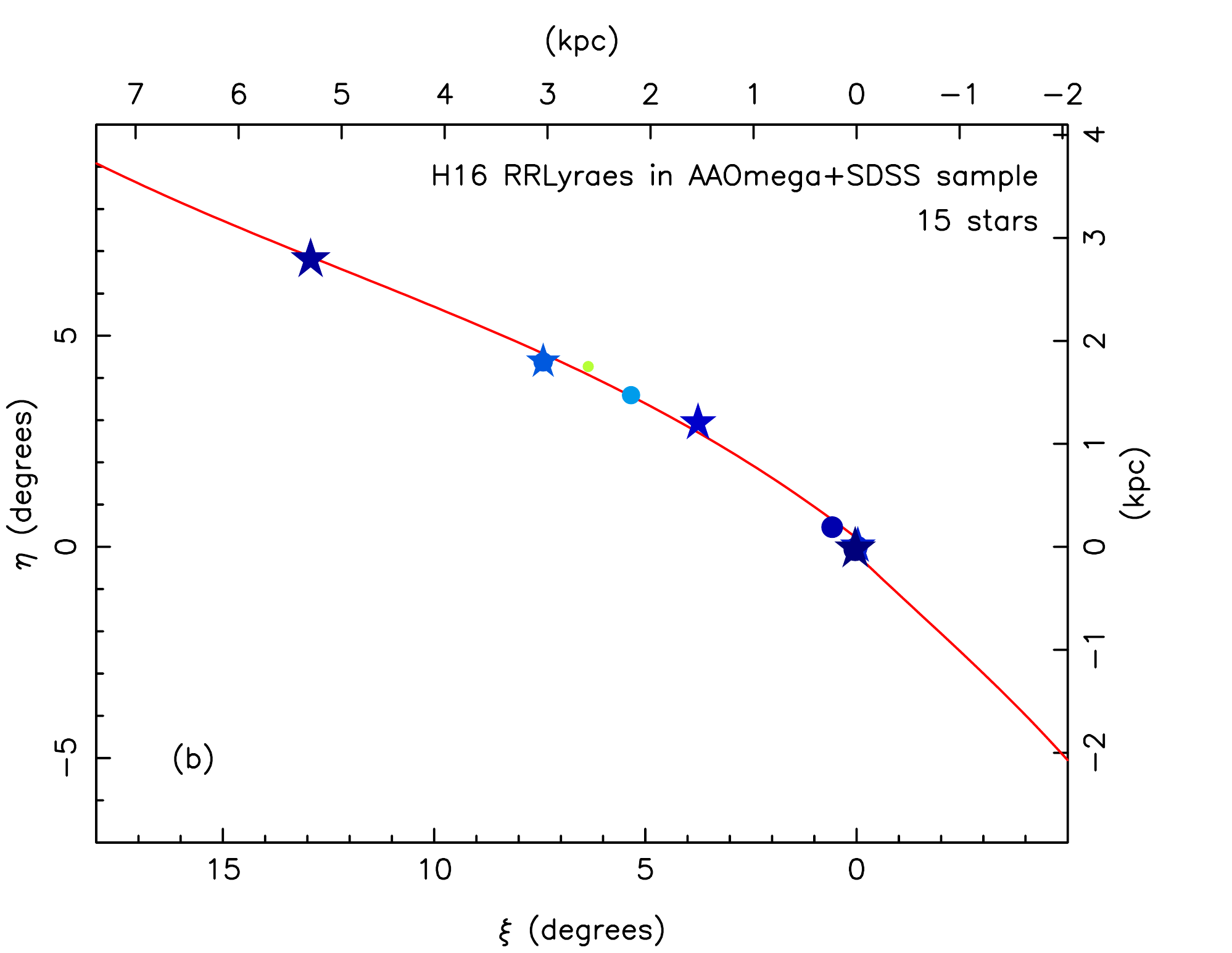}
}}
\end{center}
\caption{As Figure~\ref{fig:sel_A}, but for the RRLyrae candidate stars from the H16 catalogue, selected in the manner described in Figure~\ref{fig:RRLyrae1}, and that also have a measured radial velocity in either our AAOmega survey (circles) or in the SDSS DR12 (star symbols). As in Figure~\ref{fig:RRLyrae1}, the color and size of the graph markers denote RRLyrae probability. We note that the star near $\xi \sim 7\degg5$ was observed by us and by the SDSS at different epochs; the large velocity difference is not unexpected if this object is indeed an RRLyrae variable. Indeed, bearing in  mind the velocity variability of these objects, it is clear that all of these RRLyrae candidate stars fall in, or close to, the velocity fit from Figure~\ref{fig:sel_A}. }
\label{fig:RRLyrae2}
\end{figure}

Some additional constraints can be gained by re-examining the 22 stars in the \citet{Odenkirchen:2009js} sample given the new photometric measurements presented in Paper~I. In Figure~\ref{fig:sel_Oden}, we display their data, along with a membership probability, defined as $P_{\rm membership}=P_{\rm RGB} \times P_{\rm DDO 51}$ (the BS12 and metallicity criteria have to be ignored, as it is beyond the scope of the present work to re-analyse their raw spectra). Selecting stars, as before, from within $\pm 15\kms$ of the straight line fit previously shown in Figure~\ref{fig:sel_A}a (dashed line), we obtain the sky distribution of Figure~\ref{fig:sel_Oden}b. The higher probability RGB stars with $P_{\rm membership}>0.2$ all lie extremely close to the stream positional fit from Paper~I, and we see that the only discrepant star (near $\xi=4\degg5$) is actually a probable interloper.

A very useful additional sample for the present study comes from the analysis of probable RRLyrae stars from PanSTARRS light-curves \citep[hereafter H16]{2016ApJ...817...73H}. In Figure~\ref{fig:RRLyrae1}, we show the distances (panel a) and positions (b) of plausible RRLyrae stars ($P_{\rm RRLyrae}>0.1$, according to the H16 analysis) that lie within $0\degg3$ (i.e. $2\sigma$) of the positional fit from Paper~I. The size and color of the points code the RRLyrae probability, as computed by H16 from the light curves of these variable sources.

The distances are calculated assuming an absolute magnitude of $M_r=+0.6$ (as argued by H16 and \citealt{2013ApJ...776...26S}). In (a), one clearly sees a good correspondence between the distance fit from Paper~I (red line) and the distances derived from the mean magnitudes of the candidate RRLyraes, especially the high probability candidates. Those stars that deviate by more than $5\kpc$ from the red line have been marked as likely non-members (open symbols). It is interesting to note here that the stars below $\xi < 3\deg$ and at distances of $d_{\rm helio} \sim 50\kpc$ are most likely due to the stream of the Sagittarius dwarf galaxy, which intersects these fields (see also Paper~I).

Several of the RRLyrae candidate stars displayed in Figure~\ref{fig:RRLyrae1} have measured velocities, either from our AAOmega survey or from the Sloan Digital Sky Survey (SDSS). We show this information in Figure~\ref{fig:RRLyrae2}. Although there is considerable scatter about the straight line fit to the velocity gradient of the stream (measured in Figure~\ref{fig:sel_A}), such scatter should be expected given that RRLyraes are also radial velocity variable stars (with amplitudes of $60$--$70\kms$ for RRab stars and $30$--$40\kms$ for RRc stars; \citealt{Smith:2003vc}). The fact that all these stars with velocity measurements lie close to the fitted velocity relation lends strong credence to the large extent of the \Pal\ stream in RRLyrae stars, and it also provides confirmation and corroboration that our position, distance and velocity models are correct.

\begin{figure}
\begin{center}
\includegraphics[angle=0, viewport= 200 170 710 580, clip, width=\hsize]{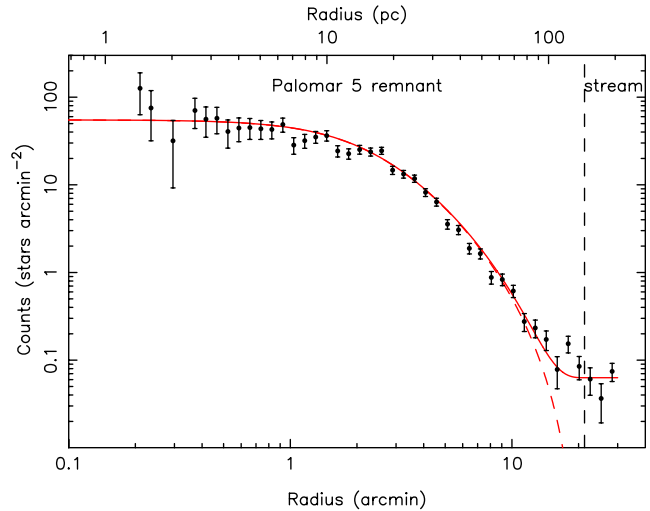}
\end{center}
\caption{Stellar number density profile. The CFHT number counts in the magnitude range ${\rm 19<g_0<23}$ and that lie within the cluster CMD selection region chosen in Paper~I, are shown as a function of radius from the cluster centre. The King model plotted over the data (dashed red line) corresponds to the most likely fit shown in Figure~\ref{fig:corner}, and has a central dispersion parameter of $\sigma_0=0.403\kms$, a tidal radius of $r_t=0.145\kpc$ ($21\mcnd2$, marked with a vertical dashed line), and a  mass of $M=4293 \msun$.The continuous red line shows the profile including the fitted contamination fraction ($C=0.111$).}
\label{fig:Profile}
\end{figure}

\section{Structural Constraints}
\label{sec:Structure}

The success of our efforts to model this system will depend sensitively on the accuracy of the parameters we use to constrain it. For this reason we next use our state-of-the-art CFHT photometry to  review the structure of the remnant, and measure the mass fraction in the tidal tails, as well as the total mass in the bound and unbound components.

\subsection{Profile of remnant}

The star number-density profile of \Pal\ can be readily measured from the CFHT catalog\footnote{We note in passing that we re-reduced the central CFHT field on the cluster using the PSF-fitting software DAOPHOT/ALLFRAME \citep{Stetson:1987fx,Stetson:1994ez}, since we were concerned that crowding in the central parts of the cluster could affect the profile there. However, it transpired that the profile from the ALLFRAME measurements was not significantly different to the profile derived from the aperture photometry measurements presented in Paper~I, so for simplicity we present the data from the Paper~I catalog.}. We chose to select stars in the magnitude range ${\rm 19<g_0<23}$, where the contrast over the foreground/background populations is best, and where the completeness is above 85\%. Figure~\ref{fig:Profile} shows the resulting background-subtracted profile. Inspection of the maps of the system (e.g. Figure 11 from Paper~I) shows that the stream emanates immediately from the cluster, and this can be seen also in the profile in Figure~\ref{fig:Profile}, where the inner King-model like structure shows a break between 10$\arcmin$--20$\arcmin$. It can be seen also from this diagram that the outer boundary of the progenitor of the stream lies at approximately $20\arcmin$.

\subsection{Stellar mass}
\label{sec:Stellar_mass}

Within this $20\arcmin$ radius region, we count the stars in the range ${\rm 19<g_0<23}$, and use  Dartmouth stellar evolution models \citep{2008ApJS..178...89D}, which can account for CFHT filter transmission curves, to correct for the missing stars. The ficucial model we use has metallicity ${\rm [Fe/H]=-1.3}$ and ${\rm [\alpha/Fe]=+0.2}$, to be consistent with the high-resolution spectroscopic study of \citet{2002AJ....123.1502S}; with these values \citet{2002AAS...201.0711M} and \citet{2011ApJ...738...74D} find a plausible age of $11.5$--$12\Gyr$ for this cluster. Assuming further a Salpeter mass function, we deduce that the stellar mass inside of $20\arcmin$ is $16700\pm 400\msun$. If instead we choose an age of $10.5\Gyr$ or $12.5\Gyr$, we derive $15000\pm 350\msun$ and $16200\pm 400\msun$, respectively. Alternatively, fixing the age at $11.5\Gyr$, but choosing ${\rm [Fe/H]=-1.2}$ and $-1.4$, gives $16600\pm 400\msun$ and $16000\pm 400\msun$, respectively. Hence the derived mass is not very sensitive to reasonable uncertainties in age or metallicity.

However, in an HST study of this system, \citet{Grillmair:2001ib} find a very shallow main-sequence slope, consistent with $d N / d m \propto m^{-0.5}$, suggesting that low-mass stars have been preferentially lost to the progenitor. With this mass function, and our fiducial parameters above (age of $11.5\Gyr$, ${\rm [Fe/H]=-1.3}$ and ${\rm [\alpha/Fe]=+0.2}$), we derive a very modest mass of $4300\pm100\msun$.

\begin{figure}
\begin{center}
\includegraphics[angle=270, viewport= 20 130 590 700, clip, width=\hsize]{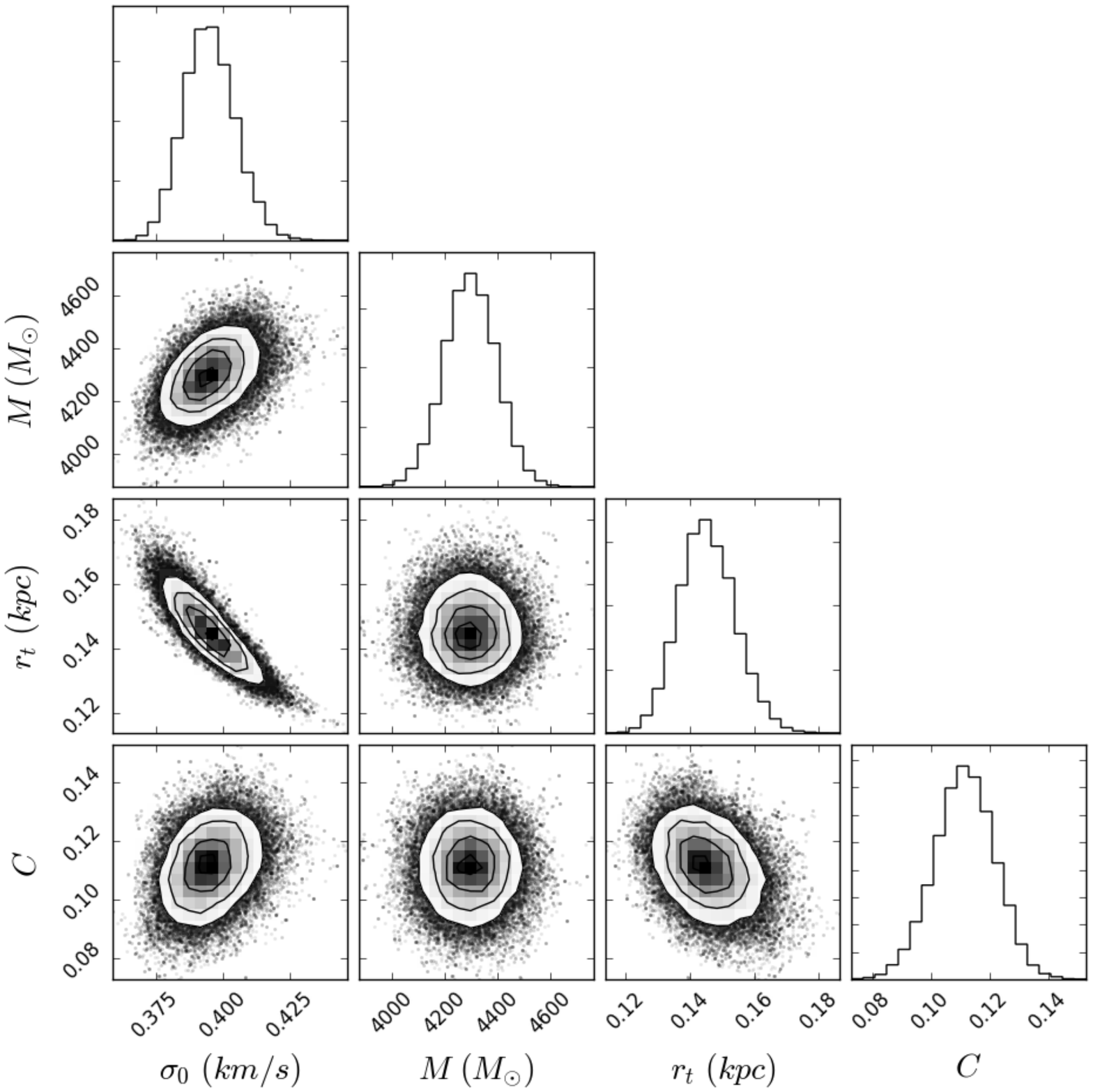}
\end{center}
\caption{An MCMC sampling of the solution space of King model fits to the \Pal\ density profile shown in Figure~\ref{fig:Profile}. The fit samples the King model parameters $\sigma_0$, the total mass $M$ and the tidal radius $r_t$, also allowing for a contamination fraction $C$.}
\label{fig:corner}
\end{figure}

\begin{figure*}
\begin{center}
\includegraphics[angle=0, viewport= 15 55 750 300, clip, width=\hsize]{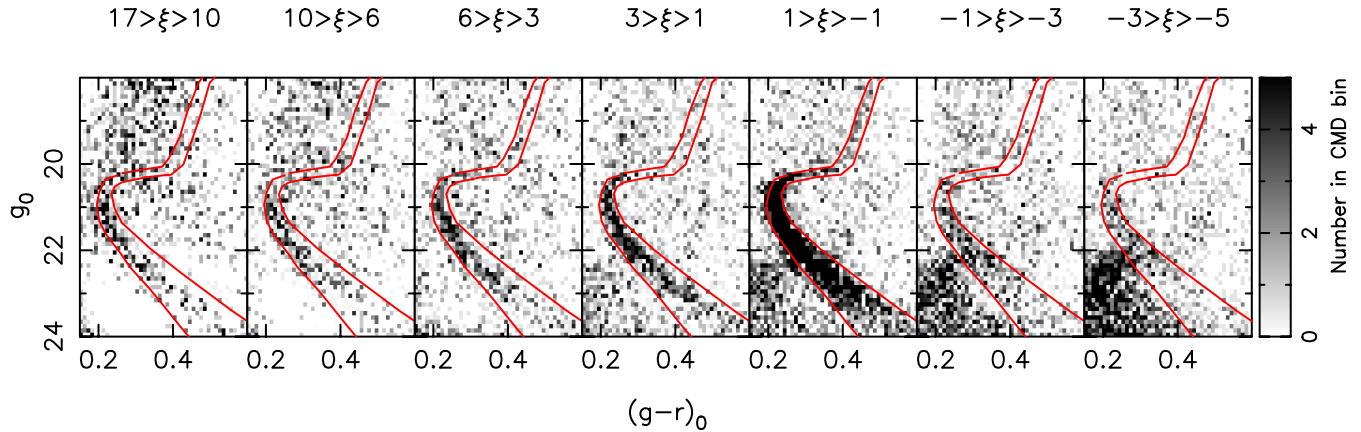}
\end{center}
\caption{Hess diagrams along the \Pal\ stream. Over each panel we mark the corresponding spatial selection region. These expanded versions of Figure~9 from Paper~I show visually that there is a dearth of  stars with ${\rm g_0>22}$ contained within the red CMD selection polygon. This is striking evidence of the lack of low-mass stars in the system. Furthermore, this property of the stellar population can be seen all along the stream. Note the presence of the main sequence turnoff of the stream of the Sagittarius dwarf galaxy at ${\rm (g-r)_0 \sim 0.2}$, ${\rm g_0>22.5}$ in those fields at $\xi \simlt 1$.}
\label{fig:Hess}
\end{figure*}

\begin{figure}
\begin{center}
\includegraphics[angle=0, viewport= 200 170 710 580, clip, width=\hsize]{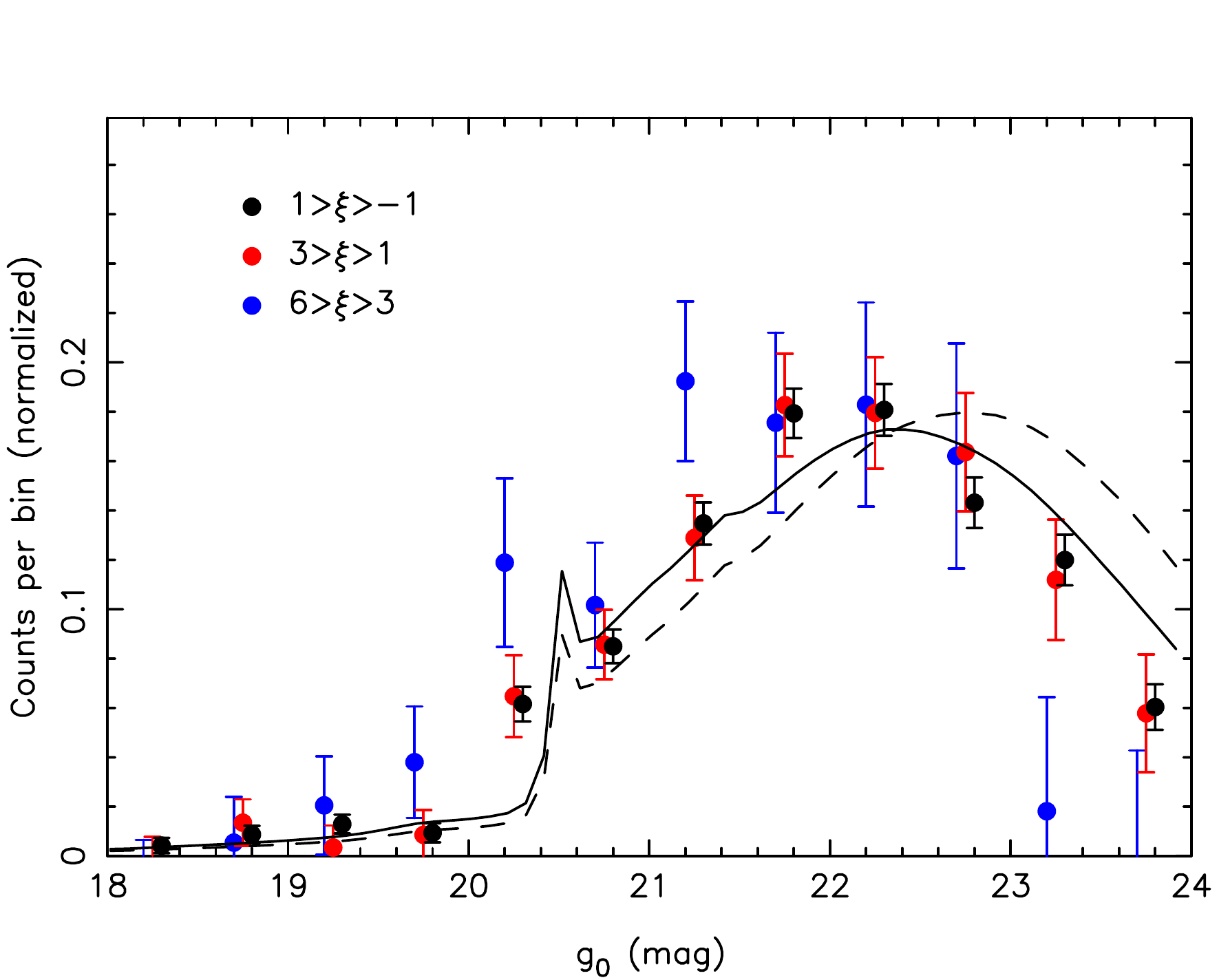}
\end{center}
\caption{The luminosity function in the central field (black circles), and in two stream regions (red and blue circles). These luminosity functions are strikingly similar, indicating that any dynamical processes that could have affected the stellar populations took place before the stars up to $\xi=6\deg$ were ejected from the cluster. For comparison, we show Dartmouth stellar population models for age $11.5\Gyr$, ${\rm [Fe/H]=-1.3}$, ${\rm [\alpha/Fe]=+0.2}$. The solid line uses a mass function $d N / d m \propto m^{-0.5}$, whereas the dashed line has a Salpeter slope.}
\label{fig:LF}
\end{figure}

This last value is consistent with the mass estimates based on SDSS data \citep{2003AJ....126.2385O}, which took into account the shallow mass function. The CFHT luminosity function of the cluster also shows a clear paucity of faint (${\rm g_0>22}$) low-mass stars, but given the existing deeper HST analysis of \citealt{Grillmair:2001ib}, the CFHT data have little to contribute to the question of the central mass, and one is forced to admit that the low mass estimates of $4300\pm100\msun$ are to be preferred. 

\subsection{Profile fits}

We need a good representation of the present mass distribution in the system to inform the dynamical modelling that we will present in the next paper in this series. To this end, we ran a set of $10^5$ fits to to the star-counts profile using a Markov Chain Monte Carlo (MCMC) algorithm. The cluster was represented as a \citet{King:1966id} model, described by a central velocity dispersion parameter $\sigma_0$, total mass $M$, tidal radius $r_t$, and we also allow the algorithm to fit for foreground and background contamination, which is included as a simple contamination fraction $C$. Figure~\ref{fig:corner} shows the solutions, from which we derive $\sigma_0=0.395\pm0.009$, $M=4297\pm98 \msun$, $r_t=0.145\pm0.009$ and $C=0.11\pm0.01$ (but note the strong parameter correlations visible in the figure). For these fits we assume ${\rm [Fe/H]=-1.3}$ and ${\rm [\alpha/Fe]=+0.2}$, and most critically we adopt the \citet{Grillmair:2001ib} mass function slope (for which there is unfortunately no uncertainty available). The most likely model solution has been overlaid on the observed profile in Figure~\ref{fig:Profile}.

It is striking that the projected central velocity dispersion of the most likely model is $0.21\kms$, very much lower than the value of $0.9\pm0.2\kms$ measured by \citet{2002AJ....124.1497O} for the whole cluster from high-resolution spectra. This difference may be due to the presence of binaries in the system, although this possibility has never been tested. 

It is worth pointing out that the observed high velocity dispersion is also problematic if a standard mass function is assumed. To explore this possibility, we ignored the stellar mass constraint presented in Section~\ref{sec:Stellar_mass} and instead took the \citet{2002AJ....124.1497O} (global) line of sight velocity dispersion as an additional datum to be fit by the above MCMC procedure. Fitting again the stellar number density profile, we found in this case a lower limit to the present day cluster mass of $M=28000\msun$ (with 99\% confidence). This is well above the allowed mass found in Section~\ref{sec:Stellar_mass} even with a Salpeter mass function slope. So the effect of binaries, or a departure from equilibrium (or both) would be required also in this case to inflate the velocity dispersion up to the observed value.

\subsection{Stellar populations variations}

The CFHT data are particularly useful in probing spatial variations in the stellar populations. That these do not change much can be perceived visually from the Hess diagrams displayed in Figure~\ref{fig:Hess}. Here the red lines mark the cluster CMD selection box, as presented in Paper~I. Within these polygons there is a clear main-sequence turnoff (which can be seen especially well in the cluster remnant region -- the fifth panel), but surprisingly, the main sequence becomes much less pronounced at $g_0>22$, despite the good completeness exceeding 70\% down to ${\rm g_0=24}$. This shows qualitatively the evidence for the scarceness  of low-mass stars mentioned in the previous section.

From those Hess diagrams, we derive the luminosity functions (LFs) shown in Figure~\ref{fig:LF}. We note that the stream at $\xi>6\deg$ is too tenuous to derive a good LF from, while the regions at $\xi<-1\deg$ are heavily contaminated by stars from the Sagittarius stream (see caption to Figure~\ref{fig:Hess}). Hence in Figure~\ref{fig:LF} we show the LF of the central field $1\deg>\xi>-1\deg$ as well as the two stream fields $3\deg>\xi>1\deg$ and $6\deg>\xi>3\deg$. The LFs of the central field (black circles) and the $3\deg>\xi>1\deg$ field (red circles) are nearly identical, and the $6\deg>\xi>3\deg$ field (blue circles) is also similar, despite the poorer statistics.

For comparison, we also overlay the Dartmouth models for stellar populations of age $11.5\Gyr$, metallicity ${\rm [Fe/H]=-1.3}$, and alpha-abundance ${\rm [\alpha/Fe]=+0.2}$. The continuous line uses a mass function of slope $-0.5$, as found by \citet{Grillmair:2001ib}, whereas the dashed line is for a Salpeter slope. While these data are not as deep as the HST photometry of \citet{Grillmair:2001ib}, they also clearly favor the shallower slope model.

These findings demonstrate that the absence of low mass stars in the \Pal\ stellar population is not a property that is just confined to the remnant. Indeed, a $\chi^2$-test indicates that the hypothesis that the three luminosity functions displayed in Figure~\ref{fig:LF} are identical cannot be rejected with better than 23\% probability. There is thus no evidence for any change in the stellar populations out to $\xi=6\deg$, $\sim 3\kpc$ from the cluster. This portion of the trailing arm contains 73\% of the trailing arm stars detected in the matched-filter map shown in Figure~11 of Paper~I. Stars that now inhabit these distant regions were disrupted several Gyr ago \citep{Dehnen:2004ez}. This means that whatever caused the loss of low-mass stars acted much earlier in the evolution of \Pal, and also --- very interestingly --- this cluster has remained quasi-stable {\it with a deficit of low mass stars} for a very long time.

\subsection{Fractional mass in trailing arm}

The matched filter map of \Pal\ and its stream presented in our previous contribution (Paper~I, Figure~11), provides a convenient way to probe the relative numbers of stars in the remaining progenitor versus its tidal tails. Unfortunately, the CFHT survey does not fully cover the leading arm, so we will examine the trailing arm and for the time being we will assume that the disruption occurred symmetrically. The ratio of the matched-filter counts between the trailing arm ($R>20\arcmin$) and the center ($R\le20\arcmin$) is found to be $0.91\pm0.03$. Hence, if the tidal disruption proceeded as expected in a symmetric manner to populate equally the leading and trailing arms, the total system would have had $2.83\pm0.06$ times more stars (of the type admitted by the adopted matched-filter) than those currently found within $20\arcmin$ of \Pal. Given that the highest CMD contrast lies at the main-sequence turnoff between ${\rm 20 \simlt g_0 \simlt 22}$, it is primarily those stars that are counted by the matched filter algorithm. This value should be considered a lower limit, since some trailing arm stars may still lie further along the orbit than we have as yet been able to survey, or  may have been scattered into other orbits by interactions with $\Lambda CDM$ sub-halos or Giant molecular clouds \citep{2016MNRAS.463L..17A}.

Our estimate can be checked for consistency against the counts of RRLyrae candidates. Selecting objects with $P_{\rm RRLyrae}>0.1$ from the H16 catalog, we obtain 6 within $20\arcmin$ of \Pal, and $12$ along the trailing arm from Figure~\ref{fig:RRLyrae1}b. This seems in reasonable agreement with our matched-filter estimate, especially given that the area covered by the trailing arm is considerably larger than that used to select the core sample, and hence is much more likely to suffer from foreground/background contamination.

\subsection{Initial system mass}

Combining the results from the previous sub-sections, we place a strong lower limit on the total initial mass of $12200\pm400\msun$. This takes the conservative assumption that the present day mass function of the remnant is valid throughout all of the tidal tails. If instead, the initial system had a Salpeter mass function, and dynamical evolution caused the preferential ejection of low-mass stars, the total initial mass would have been a much higher $47000\pm1500\msun$. While this latter value probably serves well as a firm upper limit to the initial mass, the current location of the majority of the low-mass stars that once made up this cluster remains a mystery. These bounds encompass the initial mass of $2\times 10^4\msun$ of the best model fit by \citet{Dehnen:2004ez} to SDSS data within $4\deg$ of the cluster.

\subsection{Leading arm - trailing arm asymmetry}

A visually striking aspect of the matched-filter map presented in Paper~I (Figure 11 of that contribution) is that the trailing arm of the cluster appears to be better populated than the leading arm, and we now know from the PanSTARRS maps presented in \citet{2016MNRAS.463.1759B} that the leading arm ends at $\xi=-6\deg$, just $\sim 1\degg5$ beyond the end of our CFHT survey. To quantify this possible asymmetry, we summed the matched-filter starcounts in the stream, but cut them at $g_0<22$, so as to avoid the contamination from the Sagittarius stream in the leading arm fields.

Ignoring, as before, the central $20\arcmin$ around the center of \Pal, we find a ratio of matched-filter counts in the leading to trailing arm of $0.60\pm0.04$. From the PanSTARRS matched-filter map, the contrast of the \Pal\ stream is low near $\xi=-6\deg$, but the structure appears relatively smooth. We therefore correct for the missing area, simply by assuming a constant density; this leads to a corrected ratio of $0.72\pm0.04$.

This measurement should be treated with some caution, however. First, there is uncertainty in the large-scale photometric calibration of our CFHT survey given that it was tied to the SDSS and these fields lie at a southern boundary of that survey. A second concern is that the contamination varies significantly along the large surveyed area of sky, yet the CFHT fields did not probe enough distance perpendicular to the stream to allow for a good spatially-varying background model to be constructed. 

Bearing these caveats in mind, it appears that the leading arm is indeed less populated than the trailing arm. The origin of this strong leading arm - trailing arm asymmetry is unknown, but it is almost certainly fossil evidence that different gravitational forces have acted on the opposing arms.

\section{Conclusions}
\label{sec:Conclusions}

We have presented a large spectroscopic survey of the stellar tidal stream formed by the disruption of the \Pal\ globular cluster. This tenuous structure is of extremely low contrast over the foreground and background stellar populations of the Milky Way, and has proven to be challenging to reveal with present instrumentation. While the contrast is good at the \Pal\ main-sequence turnoff, the corresponding stars are faint (${\rm g \sim 21}$), and very hard to measure accurate velocities for. This forced us to adopt the approach of surveying the RGB, at the cost of admitting a very large number of interlopers in the sample. 

Using a combination of medium-band DDO~51 photometry and accurate CFHT (g,r) photometry, together with proper motions, a gravity sensitive line (\ion{Mg}{1} $8807\AA$) and \ion{Ca}{2} metallicities, we were able to construct a clean sample of bona-fide \Pal\ stream stars. We find that this population follows an extremely well-defined narrow path on the sky, with common distance and velocity properties. The radial velocity profile can be closely fit by a straight line over the surveyed region, and displays a gradient of $0.699 \kms \, {\rm degree^{-1}}$ (in standard coordinate $\xi$) over $\sim 16\deg$ ($\sim 6.5\kpc$) across the sky.

We use these properties of the stream population to investigate its presence in the PanSTARRS RRLyrae candidate catalog of H16. We find a strong signal therein, showing good agreement in distance and velocity with respect to the RGB stars.

Earlier work by \citep{Grillmair:2001ib} has shown that the cluster remnant is surprisingly poor in low-mass stars, probably due to their being ejected through internal dynamical processes. Here we find a similar deficit all along the trailing arm stream out to $6\deg$ from the cluster (a region encompassing 73\% of stream stars), and we show that the stellar populations are statistically identical over this region (we do not test the leading arm due to the contaminating presence of the Sagittarius stream in those fields). This means that whatever process ejected the low mass stars from \Pal\ acted before the bulk of the stream stars were tidally removed from the system.

We present a measurement of the stellar number density profile in the remnant, with much better statistics than what has been derived from earlier work. We find that the remnant can be well reproduced by a King model with total mass $M=4297\pm98 \msun$, tidal radius  $r_t=0.145\pm0.009\kpc$ and central dispersion parameter $\sigma_0=0.395\pm0.009$. The corresponding projected velocity dispersion of this model is much lower than observed \citep{2002AJ....124.1497O}. An interesting avenue for future work will be to monitor temporally the velocities of stars in this cluster to ascertain whether the single epoch velocity dispersions are greatly inflated due to the presence of binaries, or whether dark matter may be present in this system. 

By correcting for the number of stars that have been lost along the stream, we estimate the pre-disruption mass to be $12200\pm400\msun$, yet the initial mass may have been as high as $47000\pm1500\msun$, if the initial mass function had a Salpeter slope. 

The next contribution in this series will present the dynamical modelling of these observations.

\section*{Acknowledgements}

We thank the anonymous referee for very helpful suggestions that have improved this paper.

Based in part on observations at Kitt Peak National Observatory, National Optical Astronomy Observatory (NOAO Prop. ID: 2010A-0088; PI: Ibata), which is operated by the Association of Universities for Research in Astronomy (AURA) under cooperative agreement with the National Science Foundation. The authors are honored to be permitted to conduct astronomical research on Iolkam Du'ag (Kitt Peak), a mountain with particular significance to the Tohono O'odham. 

\bibliography{ms}
\bibliographystyle{apj}

\begin{table}
\caption{The first 10 rows of the kinematic catalog.}
\label{tab:kinematics}
\begin{tabular}{cccccccc}
\hline
\hline
$\alpha$ & $\delta$          & $\rm{g_0}$ & $\rm{(g-r)_0}$ & $v_{helio}$    & $\delta{v_{helio}}$ & $\rm{P}$ & Inst.\\
$(^{h~m~s})$ & $(\deg~'~'')$ &  (mag)  & (mag) & ($\kms$) & ($\kms$)      &          & \\
\hline
15:03:27.723 & -03:25:23.50 & 19.130 &  0.431 &  -68.51 &  3.21 &  0.224 &F \\
15:05:55.371 & -02:50:42.50 & 17.798 &  0.482 &   14.22 &  4.33 &  0.315 &F \\
15:06:49.473 & -02:17:51.60 & 18.091 &  0.462 & -121.45 &  4.61 &  0.272 &F \\
15:08:10.594 & -02:10:07.40 & 19.161 &  0.437 & -185.13 &  2.47 &  0.601 &F \\
15:08:07.141 & -02:06:38.80 & 17.430 &  0.548 &  -59.48 &  1.11 &  0.376 &F \\
15:08:18.902 & -02:12:56.40 & 18.021 &  0.494 &  -62.94 &  0.69 &  0.922 &F \\
15:08:42.941 & -01:49:03.20 & 18.631 &  0.473 &  -56.76 &  2.40 &  0.380 &F \\
15:11:21.711 & -01:29:02.20 & 17.631 &  0.537 &  -53.06 &  0.82 &  0.965 &F \\
15:10:51.488 & -01:31:39.10 & 18.803 &  0.466 &  -58.64 &  2.00 &  0.449 &F \\
15:10:39.039 & -01:26:45.20 & 16.969 &  0.573 &  -57.04 &  1.15 &  0.964 &F \\
\hline
\hline
\end{tabular}
\tablecomments{$\alpha$ and $\delta$ list the position of the star. The 3$^{\rm rd}$ and 4$^{\rm th}$ columns list their magnitude and color. The Heliocentric radial velocity and corresponding uncertainty are given in columns 5 and 6. The membership probability is listed in column 7, while column 8 gives the instrument used to measure the star (`F' for FLAMES, `A' for AAOmega).}
\end{table}

\end{document}